\documentclass[fleqn,usenatbib]{mnras}

\usepackage{newtxtext,newtxmath}
\usepackage[T1]{fontenc}
\usepackage{ae,aecompl}

\usepackage{graphicx}	% Including figure files
\usepackage{amsmath}	% Advanced maths commands
\usepackage{amssymb}	% Extra maths symbols
\usepackage{threeparttable}
\usepackage{booktabs}
\usepackage[export]{adjustbox}
\usepackage[dvipsnames]{xcolor}
\usepackage[normalem]{ulem}
\usepackage{soul}
\newcommand{\Ion}[2]{#1\,{\sc #2}}
\newcommand{\kms}{km s$^{-1}$}
\newcommand{\Ha}{H$\alpha$}
\newcommand{\Hb}{H$\beta$}

\newcommand{\Line}[3]{#1\,{\sc #2}~$\lambda$#3}
\newcommand{\he}[1] {He\,{\sc #1}}
\newcommand{\hel}[2] {He\,{\sc #1}~$\lambda$#2}
\newcommand{\Msun}{\mbox{$\mathrm{M}_{\rm \odot}$}}
\newcommand{\Rsun}{\mbox{$\mathrm{R}_{\rm \odot}$}}
\newcommand{\target}{KR~Aur}
\newcommand{\ergs}{erg\,s$^{-1}$\,cm$^{-2}$}

\title[Dynamical masses of KR Aur in the low state]{When the disc's away, the stars will play: dynamical masses in the nova-like variable KR Aur with a pinch of accretion}

\author[P. Rodr\'\i guez-Gil et al.]{P. Rodr\'\i guez-Gil$^{1,2}$\thanks{E-mail:
prguez@iac.es}, T. Shahbaz$^{1, 2}$, M. A. P. Torres$^{1,2,3}$,
B. T. G{\"a}nsicke$^{4}$,
\and
P. Izquierdo$^{1,2}$, O. Toloza$^{4}$, A. \'Alvarez-Hern\'andez$^{1, 2}$, D. Steeghs$^{4}$, 
\and
L. van Spaandonk$^{4,5}$, D. Koester$^{6}$, D. Rodr\'\i guez$^{7}$\\ 
$^{1}$Instituto de Astrof\'\i sica de Canarias, V\'\i a L\'actea s/n, E-38205, La Laguna, Tenerife, Spain \\
$^{2}$Departamento de Astrof\'\i sica, Universidad de La Laguna, E-38206, La Laguna, Tenerife, Spain \\
$^{3}$SRON, Netherlands Institute for Space Research, Sorbonnelaan 2, 3584 CA, Utrecht, The Netherlands \\
$^{4}$Department of Physics, University of Warwick, Coventry CV4 7AL, UK \\
$^{5}$Mollerlyceum, 4611DX, Bergen op Zoom, The Netherlands \\
$^{6}$Institut f\"ur Theoretische Physik und Astrophysik, University of Kiel, D-24098 Kiel, Germany \\
$^{7}$Observatorio Guadarrama (MPC 458), E-28430 Madrid, Spain}

\date{Accepted XXX. Received YYY; in original form ZZZ}

\pubyear{2019}

\begin{document}
\label{firstpage}
\pagerange{\pageref{firstpage}--\pageref{lastpage}}
\maketitle

%%% ABSTRACT %%%

\begin{abstract}
We obtained time-resolved optical photometry and spectroscopy of the nova-like variable KR\,Aurigae in the low state. The spectrum reveals a DAB white dwarf and a mid-M dwarf companion. Using the companion star's $i$-band ellipsoidal modulation we refine the binary orbital period to be $P = 3.906519 \pm 0.000001$\,h. The light curve and the spectra show flaring activity due to episodic accretion. One of these events produced brightness oscillations at a period of 27.4 min, that we suggest to be related with the rotation period of a possibly magnetic white dwarf at either 27.4 or 54.8 min. Spectral modelling provided a spectral type of M4--5 for the companion star and $T_{1}=27\,148$\,$\pm\,496$\,K, $\log\,g=8.90 \pm 0.07$, and $\log (\mathrm{He/H})= -0.79^{+0.07}_{-0.08}$ for the white dwarf. By simultaneously fitting absorption- and emission-line radial velocity curves and the ellipsoidal light curve, we determined the stellar masses to be $M_1 = 0.94^{+0.15}_{-0.11}\,\Msun$ and $M_2 = 0.37^{+0.07}_{-0.07}\,\Msun$ for the white dwarf and the M-dwarf, respectively, and an orbital inclination of $47^{+1^{\rm o}}_{-2^{\rm o}}$. Finally, we analyse time-resolved spectroscopy acquired when the system was at an $i$-band magnitude of 17.1, about 1.3 mag brighter than it was in the low state. In this intermediate state the line profiles contain an emission S-wave delayed by $\simeq 0.2$ orbital cycle relative to the motion of the white dwarf, similar to what is observed in SW Sextantis stars in the high state.
\end{abstract}


%%%%%%%%%%%%%%%%%%%%%%%%%%%%%%%%%%%%%%%%%%%%%%%%%%

%%%%%%%%%%%%%%%%% BODY OF PAPER %%%%%%%%%%%%%%%%%%

%%% INTRODUCTION %%%

\section{Introduction} \label{sec:intro}

In cataclysmic variables (CVs) a white dwarf (WD) and a less massive, late-type companion star orbit around a common centre of mass. The companion fills its Roche lobe so material is spilt into the larger potential well of the WD, and is finally accreted usually via an accretion disc. Observationally, the orbital period distribution of CVs exhibits an apparent deficiency of mass-transferring systems in the 2--3 hour range \citep[see e.g.][and references therein]{gaensickeetal09-1}. This so-called ``period-gap'' is the key feature that has been defining the current standard scenario of CV evolution, based on the concept of ``disrupted magnetic braking'' \citep{king88-1, knigge06-1}. In this framework, CVs are thought to evolve from long to short orbital periods as their orbits lose angular momentum, with magnetic braking \citep{verbunt+zwaan81-1} being the dominant driving mechanism at an orbital period $P \ge 3$\,h. The mass transfer results in the mass-donor star being driven slightly out of thermal equilibrium, becoming over-sized for its mass, and the assumption is that the magnetic braking ceases once the companion becomes fully convective at $P\simeq3$\,h. The donor star reacts by contracting to its main-sequence radius corresponding to its mass, effectively shutting down mass transfer~---the CV detaches and enters the period gap. Gravitational wave radiation continues to drive the systems to shorter periods, though on longer time scales, until their companion stars eventually re-fill their Roche lobes at $P\simeq2$\,h. 

A striking feature among the observed population of CVs is a build-up of intrinsically bright nova-like variables in the 3--4 hour interval which show a coherent behaviour, called the SW\,Sextantis stars after their prototype, initially described by \citet{thorstensenetal91-1} and later reviewed by \citet{rodriguez-giletal07-1}. Where estimated, the mass-transfer rates of these systems are very high, up to $\simeq10^{-8}\,\mathrm{M_\odot\,yr^{-1}}$  \citep{townsley+gaensicke09-1}, in fact exceeding the predictions of canonical CV evolution models (e.g. \citealt{howelletal01-1, kniggeetal11-1, kalomenietal16-1}). 

In addition, a number of nova-like variables in the 3--4\,h period range are known to alternate between states of high and low mass transfer rates. In the high state the accretion luminosity outshines both the WD and the companion star. In contrast, during the occasional low states systems fade by 3--5 magnitudes, revealing the companion star and the WD \citep[see][]{gaensickeetal99-1,hoardetal04-1,rodriguez-giletal12-1,rodriguez-giletal15-1}, and making measurement of fundamental parameters of the binary systems possible. Nova-like variables exhibiting low states are known as VY Sculptoris (VY\,Scl) stars. Low states are likely a result of a reduction in the mass transfer rate from the companion star, but the underlying physical cause is still unknown. Quenching of mass transfer caused by starspots \citep{livio+pringle94-1,king+cannizzo98-1} or the effect of the variable and active chromosphere of the companion star \citep{howelletal00-1} have been invoked.

Improving our understanding of the currently unexplained behaviour of the nova-like variables in the 3--4\,h period range is largely hampered by the lack of a significant number of systems with accurate stellar and binary parameters, requiring high-quality spectroscopic and photometric observations of the stellar components during low states. To date, only the eclipsing nova-like variable HS\,0220+0603 has dynamical masses measured \citep{rodriguez-giletal15-1}. Here, we present the second such study, a dynamical analysis of the non-eclipsing, nova-like variable KR\,Aur. The paper is organised as follows: in Section~\ref{sec:nearby} we give an introduction on \target. In Section~\ref{sec:obs} we present the observations and the data reduction process. Section~\ref{sec:pblc} shows the long-term light curve of KR\,Aur and an analysis of time-resolved photometry in the low state, obtaining a refined ephemeris. Spectral modelling of the WD and the companion star is detailed in Section~\ref{sec:spec_class}, and Section~\ref{sec:interm} gives a comparison between optical spectra during the low and the intermediate state. A radial velocity study from data in these two states is presented in Section~\ref{sec:RVCs}. The binary parameters are determined in Section~\ref{sec:binparam}. Finally, we draw our conclusions in Section~\ref{sec:conclu}.\\

\section{KR Aur, a VY Scl system}
\label{sec:nearby}

The variability of KR\,Aur ($=$ SON 5420) was discovered by \citet{popova60-1} from comparison of photographic plates, who reported it as a red object ``slowly varying'' between 12.5 and fainter than 14.5 magnitude \citep{popowa61-1}. \cite{hoffmeister65-1} later described a quite erratic long-term photometric behaviour after examining photographic plates taken sparsely in the period 1942--1965. He reported KR\,Aur as a blue object varying back and forth from $\simeq$18--19 to $\simeq$11--13 mag, which prompted him to (mis-)classify it as a nova (Nova Aur 1960). \cite{doroshenkoetal77-1} showed that the photometric and spectroscopic characteristics of KR\,Aur agreed with it being a close binary system. Later, \cite{liller80-2} confirmed the long-term variability of KR\,Aur and \cite{mufsonetal80-1} reported X-ray emission in the 0.15--4.5\,keV energy range. The CV nature of the system was finally settled by means of time-resolved spectroscopy, establishing an orbital period $P = 0.16280$\,d (\citealt{shafter83-1}, later refined to 0.16274\,d, \citealt{hutchingsetal83-1}), and a first estimate of the stellar parameters, $M_1 \sim 0.7$~\Msun, a companion star mass of $M_2 = 0.48$~\Msun, and a binary inclination of $i \leq 40^{\circ}$. 
 
\cite{biryukov+borisov90-1} reported a modulation at $\simeq25$\,min in a $B$-band light curve of KR\,Aur in the high state, that \cite{singhetal93-1} and \cite{katoetal02-2} later failed to recover. Instead, they claimed quasi-periodic variability with time scales of the order of tens of minutes and significant period changes from night to night. The same behaviour was observed in the high-state light curves obtained by \cite{kozhevnikov07-1}, who also detected a negative superhump with a period of 3.771\,h, 3.5 per cent shorter than Shafter's (\citeyear{shafter83-1}) value for the orbital period.

\begin{figure*}
\centering
\includegraphics[width=0.95\linewidth]{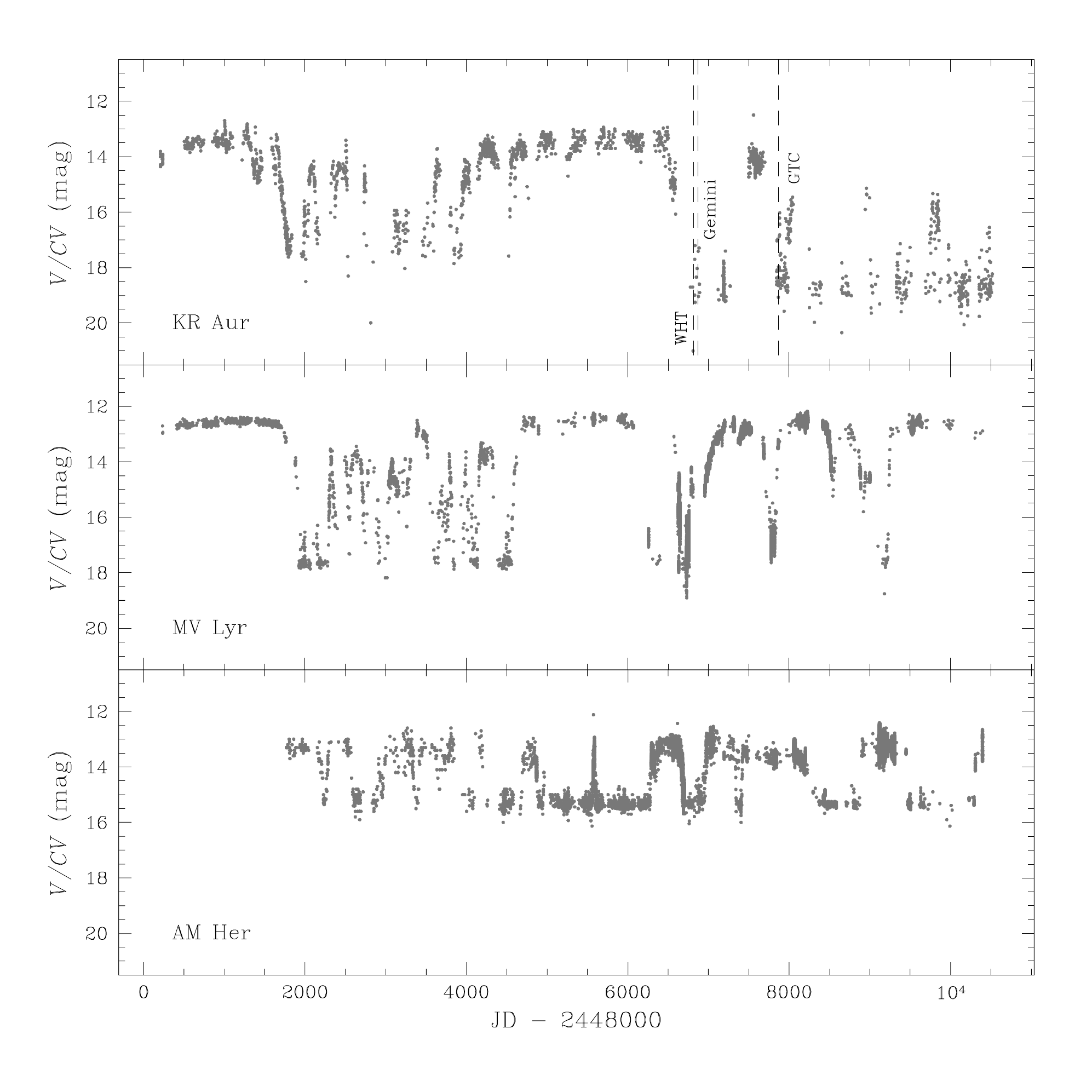}
\caption{Long-term, Clear-$V$ ($CV$) plus $V$-band light curves of KR\,Aurigae (top panel) showing 28.23 years of photometric coverage from 1990 November 12 to 2019 February 6, the nova-like variable MV Lyrae (middle panel) covering 27.82 years from 1990 December 10 to 2018 October 4, and AM Herculis (bottom panel), the prototype of the highly-magnetic CVs, from 1995 February 19 to 2018 October 4, 23.62 years of photometric coverage. Note the different high/low brightness amplitude between the systems that have an accretion disc in the high state (KR\,Aur and MV\,Lyr) and AM\,Her, which does not. All light curves from AAVSO data. Most data points during the first low state of KR\,Aur were provided by the 0.41-m RoboScope \citep{honeycutt+turner92-1}. Exposures taken with this telescope had a typical magnitude limit $V \simeq18$ under dark sky conditions. This may explain why the first recorded low state of \target\ looks shallower than the second one.}
\label{fig_longterm3}
\end{figure*}

In Fig.~\ref{fig_longterm3} we compare the long-term light curves of three CVs that show low states: the nova-like variables KR\,Aur and MV\,Lyr, as well as the prototype of the strongly magnetic polar CVs, AM\,Her. All three systems exhibit repeated, and often relatively rapid, changes between low and high states. Because of the absence of an accretion disc in AM\,Her the magnitude difference between its high and low state is smaller compared to the two nova-like CVs. Further, the lack of an accretion disc in AM\,Her points to quenching of the mass transfer rate from the companion star as an essential ingredient to produce a low state.

Finally, \cite{bailer-jonesetal18-1} derived a distance to the system from \textit{Gaia} Data Release 2 (DR2) of $451^{+112}_{-75}$\,pc, placing \target\ quite nearby.

%%% OBSERVATIONS AND REDUCTION %%%

\begin{table*}
\caption{Log of spectroscopy observations.}
\centering
\begin{threeparttable}
\begin{tabular}{@{}lccccc}
\hline
UT date &  Grating & Exposure time& $\lambda$ range & Resolution & \# of science \\
&  & (s)  & (\AA) & (\AA) & frames\\
\hline
{\textbf{WHT/ISIS-blue}~~~~~~~} & & & & & \\
2008 Dec 7 & R600B & 600 & 3760--5560 & 2.0 & 4 \\
2008 Dec 8  & R600B & 600 & 3650--5430 & 2.0 & 11 \\
2008 Dec 16 & R300B &1200 & 3640--5320 & 3.4 & 17 \\
{\textbf{WHT/ISIS-red}} & & & \\
2008 Dec 7 & R600R & 600 & 7720--9750 & 2.0 & 4 \\
2008 Dec 8 & R600R & 600 & 5310--7330 & 2.0 & 10 \\
2008 Dec 16 & R600R & 1200 & 6430--7930 & 1.6 & 17 \\
\hline
{\textbf{Gemini/GMOS}}  &  &  &  \\
2009 Feb 2 & R831 & 840 & 6345--8565 & 1.6 & 16 \\
\hline
{\textbf{GTC/OSIRIS}}  &  &  &  \\
2011 Nov 5 & R2500R & 600 & 5575--7685 & 2.0 & 14 \\
\hline
\end{tabular}
\end{threeparttable}
\label{tab:slog}
\end{table*}

\section{Observations and data reduction}
\label{sec:obs}

% SPECTROSCOPY

\subsection{Spectroscopy}
\label{sec:spec}

\subsubsection{William Herschel Telescope}

We obtained spectra of KR\,Aur with the 4.2-m William Herschel Telescope (WHT) and the double-armed Intermediate dispersion Spectrograph and Imaging System (ISIS) at the Roque de los Muchachos Observatory (La Palma) on 2008 April 22 and December 7, 8 and 16. The 4096$\times$2048 pixel EEV12 and the 4096$\times$2048 pixel RED+ CCD cameras were used in the blue and the red arm, respectively.

On December 7 and 8 we opted for the R600B (blue) and R600R (red) gratings. The GG495 second-order sorter filter was also placed in the red arm's light path. We used a 1.2-arcsec slit, central wavelengths of 4670(4540)\,\AA\ for the blue arm and 8600(6180)\,\AA\ for the red arm on December 7(8). We binned both CCD detectors by factors of 2 in both the spatial and spectral directions. The exposure time was 600\,s and the spectral resolution (full-width at half-maximum, FWHM) is about 2.0\,\AA. On December 16 we orientated the 1-arcsec slit to a sky position angle of $114^\circ$ (west from north) to accommodate a comparison star $1.07$ arcmin away from KR\,Aur and performed relative spectrophotometry. We used gratings R300B and R600R ($+$GG495 filter) and respective central wavelengths of 4500 and 7090\,\AA. Binning by a factor of 4 was applied only to the spatial direction. The exposure time was fixed to 1200\,s and FWHM spectral resolutions of 3.4 (blue) and 1.6\,\AA\ (red) were achieved. We also took spectra of the G191--B2B spectrophotometric standard on December 16 to correct for the instrumental response. In order to account for instrument flexure we regularly took spectra of CuNe+CuAr arc lamps on all the nights. See Table \ref{tab:slog} for details on the spectroscopy observations.

\subsubsection{Gemini North Telescope}

We obtained time-resolved spectroscopy of KR\,Aur at the 8.1-m Gemini North Telescope at Manua Kea, Hawaii, with the Gemini Multi-Object Spectrograph (GMOS) on 2009 February 2, as part of our target-of-opportunity (ToO) GN--2008B--Q--39 programme. The detector consists of three adjacent 2048$\times$4068 pixel CCDs which are separated by two gaps of $\simeq 37$\,pixels each. We used the R831 grating and the RG610 blocking filter, on-detector 2$\times$2 binning, and the standard nod-and-shuffle mode with a 0.5-arcsec slit. The telescope nods between position A and position B (6-arcsec offset in our case), so the sky background at the same position as the target is recorded very close in time. The exposure time was fixed at 60\,s. To obtain every individual spectrum of KR\,Aur we ran seven nodding cycles to get a total exposure time of $2\,\mathrm{(positions)}\,\times\,7\,\mathrm{(cycles)}\,\times\,60=840$\,s. The instrumental setup sampled the 6345--8565\,\AA\ range at 1.6-\AA\ FWHM resolution. In order to fill in the detector gaps we used dithering in the spectral direction by changing the central wavelength of the grating between 7400, 7450, 7500 and 7550\,\AA\ (the wavelength range quoted in Table~\ref{tab:slog} is the common range shared by all spectra). Wavelength calibration relied on CuAr arc lamp exposures taken during daytime.

\begin{table}
\centering
\caption{Log of photometry observations.}
\begin{tabular}{@{}lccc}
\hline
Date & Filter & Exp. time& Coverage \\
&  & (s) & (h) \\
\hline
{\textbf{INT/WFC}} &  &  &   \\
2009 Feb 11 & $i$ & 60 & 3.33  \\
2010 Jan 4 & $i$ & 60 & 7.04  \\
2010 Feb 9 & $i$ & 60 & 3.99  \\
2017 Feb 2  & $i$ & 60 & 3.50  \\

\hline
{\textbf{JKT/Andor CCD}} &  &  &   \\
2017 Feb 18 & $i$ & 240 & 5.95  \\

\hline
{\textbf{IAC80/CAMELOT}} &  &  &   \\
2011 Dec 23 & $i$ & 270 & 4.01  \\
2014 Feb 3 & $i$ & 180, 240 & 4.89  \\
2015 Jan 24 & $i$ & 240 & 3.77  \\
2015 Jan 25 & $i$ & 240, 320 & 3.92  \\
2015 Jan 26 & $i$ & 240 & 5.60  \\
2016 Mar 27 & $g$ & 240 & 2.88  \\
2016 Mar 31 & $g$ & 240 & 2.81  \\

\hline
\end{tabular}
\label{tab_obslog}
\end{table}

\subsubsection{Gran Telescopio Canarias}

We also observed KR\,Aur on 2011 November 5 with the 10.4-m Gran Telescopio Canarias (GTC) at the Roque de los Muchachos Observatory on La Palma. We used the Optical System for Imaging and low-Intermediate-Resolution Integrated Spectroscopy (OSIRIS) with the array of two Marconi CCD detectors, each with 2048$\times$4096 pixels. The spectra were produced with the R2500R grating and a 0.4-arcsec slit, spanning the wavelength range 5575--7685\,\AA\ at a central spectral resolution of 2.0\,\AA\ (FWHM). The exposure time was 600\,s. For wavelength calibration we used spectra of HgAr+Ne lamps obtained at the beginning of the night. The spectrophotometric standard star G191--B2B was observed on the same night. This was used to correct for the instrumental response. 

~\\

The WHT and GTC raw spectra were reduced using standard procedures in {\sc iraf}\footnote{{\sc iraf} is distributed by the National Optical Astronomy Observatories}. The Gemini observations were reduced using the \texttt{gemini} pipeline also within {\sc iraf}. Next, we did an optimal extraction of all the spectra using the {\sc starlink}/{\sc pamela} reduction package \citep{marsh89-1}. All subsequent calibration and analysis were performed with {\sc molly}\footnote{Tom Marsh's {\sc molly} package is available at \url{http://deneb.astro.warwick.ac.uk/phsaap/software}}. We finally corrected all the flux-calibrated spectra for interstellar reddening using $E(B-V)=0.07$. This value is given by the three-dimensional dust map of \cite{greenetal18-1} at the distance to KR\,Aur, $D_{\rm pc}=451$\,pc, derived from its \textit{Gaia} DR2 parallax \citep{bailer-jonesetal18-1}. The same value of the colour excess
is obtained for the above distance when employing the three-dimensional maps of \citet{lallementetal19-1}, built from \textit{Gaia}/2MASS photometry, and \textit{Gaia} DR2 parallaxes. A value $E(B-V) = 0.05 \pm 0.05$ was reported in \cite{verbunt87-1} from the interstellar feature at 2200 \AA.

% PHOTOMETRY

\subsection{Photometry}

\subsubsection{Isaac Newton Telescope}

We conducted time-resolved, $i$-band photometry of KR\,Aur with the 2.5-m Isaac Newton Telescope (INT) at the Roque de los Muchachos Observatory on La Palma. We used the Wide-Field Camera (WFC), a mosaic of four EEV 2048$\times$4096 pixel CCDs with a plate scale of 0.33 arcsec pix$^{-1}$. We only read out CCD \#4, and the exposure time was set to 60 s (108-s time resolution). See Table \ref{tab_obslog} for details.

\subsubsection{IAC80}

We also used the 0.82-m IAC80 telescope at the Observatorio del Teide on Tenerife to obtain time-resolved, $i$-band photometry of KR\,Aur. The observations were performed with the C\'amara MEjorada Ligera del Observatorio del Teide (CAMELOT) equipped with an E2V 2148$\times$2148 pixel CCD. The individual exposures ranged between 180 and 320 seconds depending on seeing conditions. Data with the $g$-band filter were also taken on 2016 March 27 and 31 (Table~\ref{tab_obslog}).

\subsubsection{JKT}

The Southeastern Association for Research in Astronomy (SARA) 1-m Jacobus Kapteyn Telescope (JKT) on La Palma \citep{keeletal17-1} provided the final $i$-band light curve of this work on 2017 Feb 18. The detector is a 2048$\times$2048 pixel Andor Ikon-L CCD with thermoelectric cooling. The exposure time was 240 seconds. 

All photometry images were reduced using standard packages within {\sc iraf}.

%%% LONG TERM LIGHT CURVE %%%

% ---------- Finding Chart ------------------------------------------------
\begin{figure}
\centering
\includegraphics[width=\columnwidth]{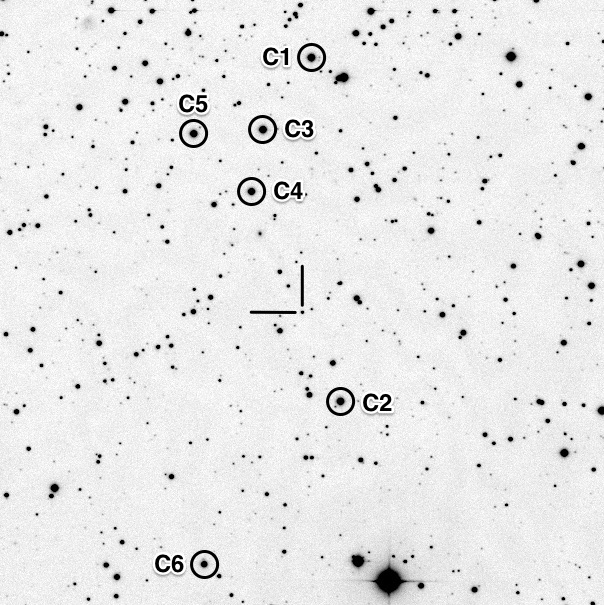}
\caption{\label{fig_chart_comps} $i$-band $7\arcmin \times 7\arcmin$ finding chart for
KR\,Aur in the low state obtained with INT/WFC on 2009 February 11. North is up, east is to the left. Magnitudes of the comparison stars C1--C6 are given in Table~\ref{tab_comps}.}
\end{figure}

%-------------------COMPARISON STARS Table--------------------------------------------
\begin{table}
\centering
\caption{\label{tab_comps} Magnitudes of the comparison stars used for the photometry (see Fig.\,\ref{fig_chart_comps}).}
\begin{threeparttable}
\begin{tabular}{cll}
\hline\noalign{\smallskip}
~~~~ID &   ~~~~~~$~~~~g$ & ~~~~~~$~~~~i$~~~ \\
\hline\noalign{\smallskip}
~~~~C1 &   ~~~~~~16.25(6) & ~~~~~~14.66(8)~~~ \\
~~~~C2 &   ~~~~~~16.13(8) & ~~~~~~14.69(10)~~~ \\
~~~~C3 &   ~~~~~~16.29(12) & ~~~~~~14.75(6)~~~ \\
~~~~C4 &   ~~~~~~16.60(7) & ~~~~~~14.84(11)~~~ \\
~~~~C5 &   ~~~~~~16.19(6) & ~~~~~~14.85(9)~~~ \\
~~~~C6 &   ~~~~~~16.40(3) & ~~~~~~15.26(2)~~~ \\
\hline\noalign{\smallskip}
\end{tabular}
    \begin{tablenotes}
    \item {\footnotesize Numbers in brackets are the uncertainties on the last digits of the preceding magnitude.}
    \end{tablenotes}
\end{threeparttable}
  \end{table}

\section{Light curve in the low state}
\label{sec:pblc}

The long-term $C$$V$/$V$-band\footnote{$C$$V$ magnitudes are computed from unfiltered images using the $V$-band magnitude of the comparison star.} AAVSO light curve of KR\,Aur (Fig.~\ref{fig_longterm3}) spans 28.23 years of photometric coverage from 1990 November 12 to 2019 February 6. KR\,Aur exhibits two distinct low-state epochs with erratic large-amplitude variability, one from 1994 October to 2001 February ($\mathrm{JD} - 2448000 \simeq 1640$ to $\simeq 3950$ in the plot), and the low state that started in 2008 November ($\mathrm{JD} - 2448000 \simeq 6780$). These low states are interspersed with well-defined high states. 

In this section we analyse time-resolved photometry obtained during the second recorded low state of KR\,Aur. To produce the light curves we performed variable-aperture photometry relative to an ensemble of six comparison stars with the AstroImageJ package \citep[{\sc aij}\footnote{\url{http://www.astro.louisville.edu/software/astroimagej/}},][]{collins+kielkopf13-1,collinsetal17-1}. The comparison stars were selected from the Data Release 9 of the AAVSO Photometric All-Sky Survey \citep[APASS,][]{hendenetal09-1} and are indicated in Fig.~\ref{fig_chart_comps}. Their APASS $g$ and $i$ magnitudes are listed in Table~\ref{tab_comps}.

\begin{figure}
\centering
\includegraphics[width=\columnwidth]{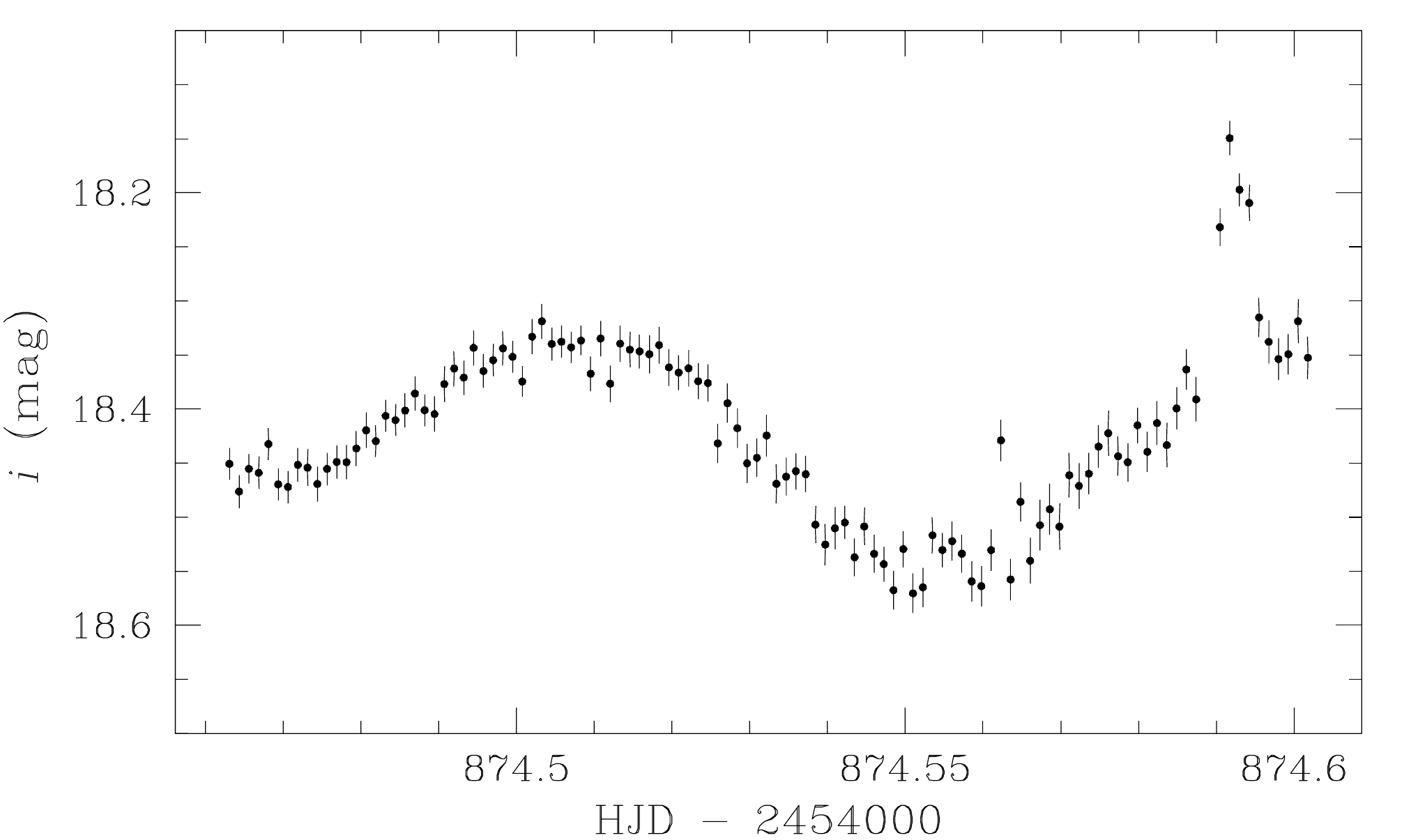}
\caption{Low-state, $i$-band light curve of KR\,Aur taken with the INT on 2009 February 11. Apart from the ellipsoidal modulation two flares are apparent.}
\label{fig_flares_20090211}
\end{figure}

The $i$-band light curve of KR\,Aur in the low state is dominated by the ellipsoidal variation caused by the changing projected area of the companion star along its orbit, with two maxima and two minima per orbital revolution. We show a sample light curve in Fig.~\ref{fig_flares_20090211}. If we adopted as phase 0.5 the time of the absolute minimum of this light curve (as expected for a canonical ellipsoidal modulation), the radial velocity curve of the \Ion{Na}{i} absorption doublet (Section~\ref{sec:RVCs}) obtained only nine days earlier would be offset by 0.5 cycle with respect to the motion of the companion star when folded on Shafter's orbital period. This 180-degree offset indicates that the deeper minimum in the ellipsoidal light curve of KR\,Aur actually takes place at inferior conjunction of the companion star (i.e. orbital phase 0), as the \Ion{Na}{i} absorption doublet originates on the companion star. A shallower minimum at phase 0.5 is consistent with the inner face of the companion star being irradiated by the WD ($T_1 \simeq 27\,000$\,K; Section~\ref{sec:spec_class}).

% ---------------------------------------------------------------------------

We observed the same behaviour in the low-state light curve of the \textit{eclipsing} SW\,Sextantis star HS\,0220+0603 \citep{rodriguez-giletal15-1}. The eclipses in this system, and hence an unambiguous definition of orbital phase 0, allowed us to confirm that the deeper minimum of the ellipsoidal variation in HS\,0220+0603 actually occurs at phase 0 instead of 0.5 due to irradiation of the inner hemisphere of the companion star by the WD. Hence, we adopt phase zero to be the mid-time of the deeper minimum in the ellipsoidal light curve of \target.

\begin{figure}
\centering
\includegraphics[width=\columnwidth]{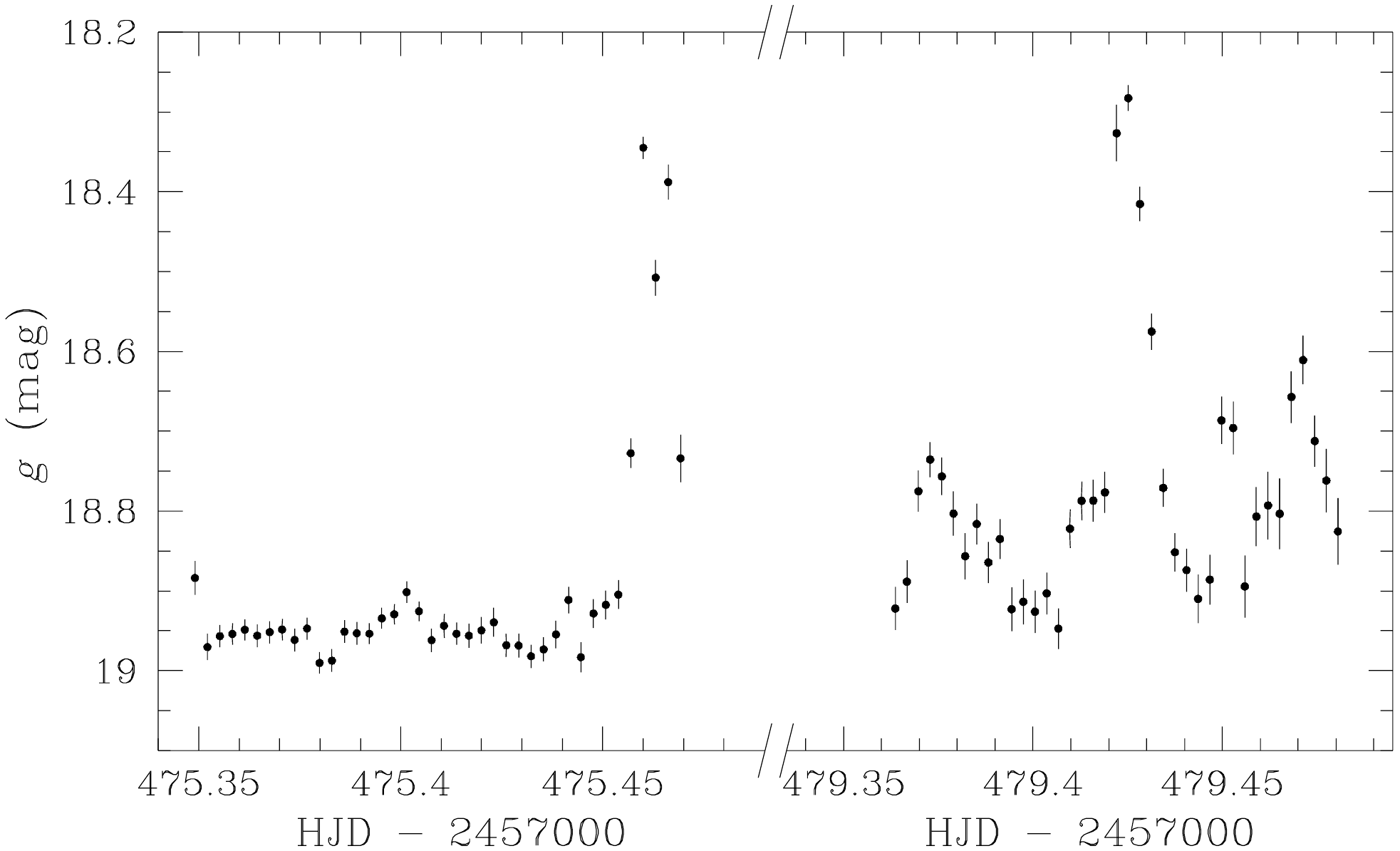}
\caption{Low-state, $g$-band light curves of KR\,Aur taken with IAC80/CAMELOT on 2016 March 27 (left) and 31 (right).}
\label{fig_flares_g}
\end{figure}

%----------------------------------------------------------------

\subsection{Flare events as accretion episodes}
\label{sec_flares}

The low-state $g$- and $i$-band light curves of KR\,Aur show flare activity (see Figs.~\ref{fig_flares_20090211} and \ref{fig_flares_g}). The flares have an amplitude of $\approx -0.5$ mag above the ellipsoidal light curve level and typical durations of 10--30 min. \cite{bonnet-bidaudetal00-1} observed similar flare activity in optical photometry obtained during the 1991 low state of the polar CV AM Her. They reported 5--10 min long flares with amplitudes ranging between $-0.24$ and $-0.67$\,mag in the $V$ band. Based on their measurement of circular polarisation in the $R$ and $I$ bands these authors attributed the flares to accretion events during the low state. \cite{kafkaetal05-2} also reported flares in the low-state, $V$-band light curve of AM Her with 15--90\,min duration and $-0.2$ to $-0.6$\,mag amplitude. The characteristics of these flares significantly differed from those of the intense, blue flare detected in AM Her on 1992 August 29 when the system was in the low state ($V=15.40$), with an amplitude of $-2.14$\,mag, a total duration of $\simeq 50$\,min in the $V$ band and unpolarised \citep{shakhovskoyetal93-1}. These authors suggested that, unlike the majority of flares observed in AM Her in the low state, this strong flare has nothing to do with accretion and was likely produced on the companion star as observed in isolated, flare M-dwarf stars.

Flaring behaviour in the low state has also been observed in CVs that possess bright accretion discs in the high state. Bursts recurring about every two hours with a duration of approximately 30\,min and amplitudes of up to $-2$\,mag have been observed in the \textit{Kepler} $V$-band light curve of the nova-like variable MV\,Lyr at $V \approx 17.5$ \citep{scaringietal17-1}. They explain these recurrent flares as caused by the magnetic gating process. The disc is truncated by the weak magnetic field of the WD ($10^{-2}-10^{-1}$\,MG), and inner-disc material orbiting \textit{outside} the co-rotation radius (the distance from the centre of the WD at which the Keplerian velocities of the disc material and the WD rotation match) is forced to accumulate by this centrifugal barrier imposed by the rotating WD magnetic field, since the WD rotates faster than the inner disc. As more matter accumulates the pressure exerted on the WD magnetic field by the inner disc increases until the barrier is overcome and accretion onto the WD takes place. This way the inner disc is depleted and the pile-up of matter around the WD magnetosphere can start again.     

\begin{figure}
\centering
\includegraphics[width=0.81\columnwidth]{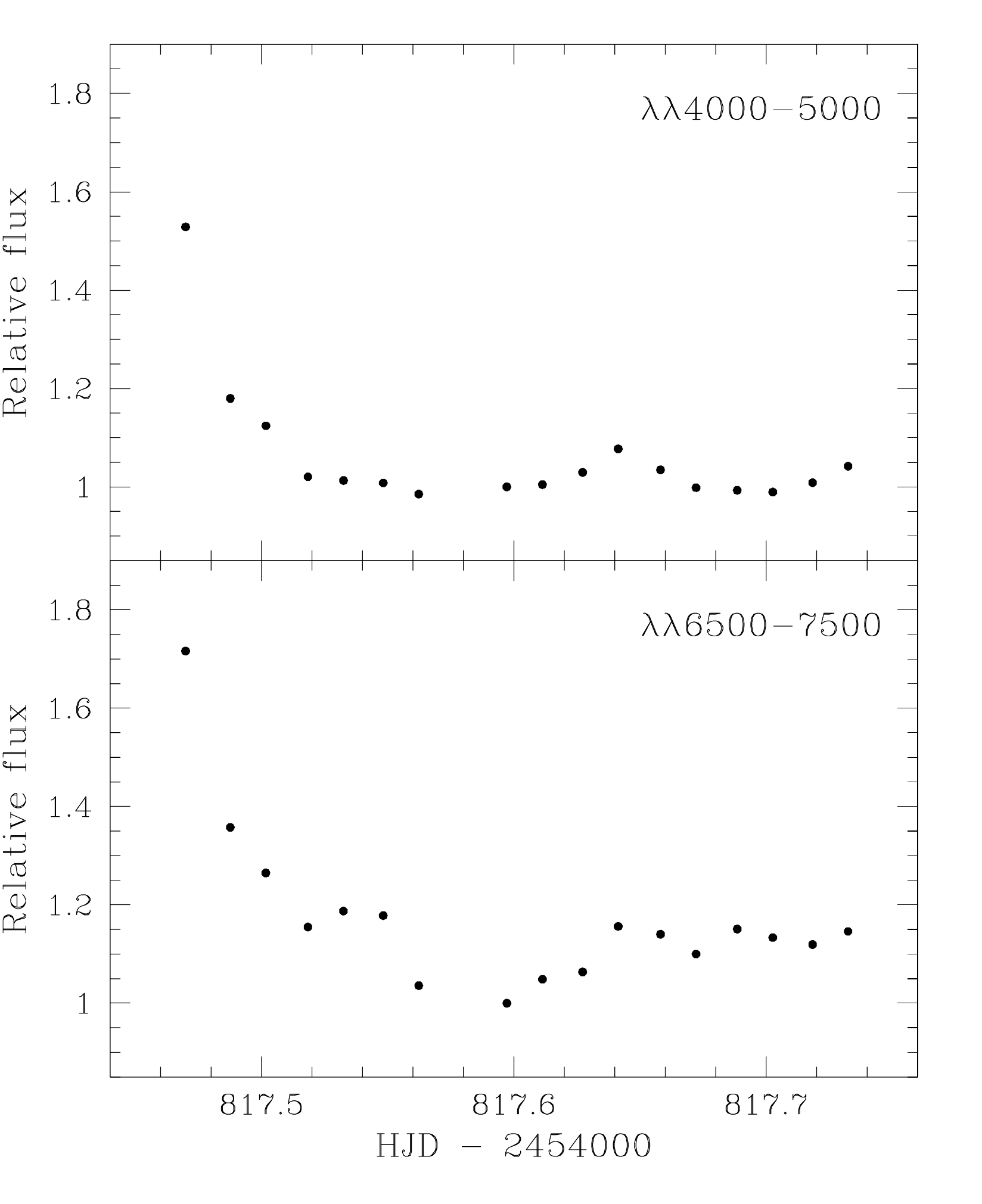}
\caption{Spectrophotometric light curves of KR\,Aur computed from the WHT/ISIS blue-arm (\textit{top}) and red-arm (\textit{bottom}) spectra obtained on 2008 December 16 relative to a nearby comparison star also on the slit. The data have been normalised to the value of spectrum \#8 of the series.}
\label{fig_flare_specphot}
\end{figure}

\begin{figure}
\centering
\includegraphics[width=\linewidth]{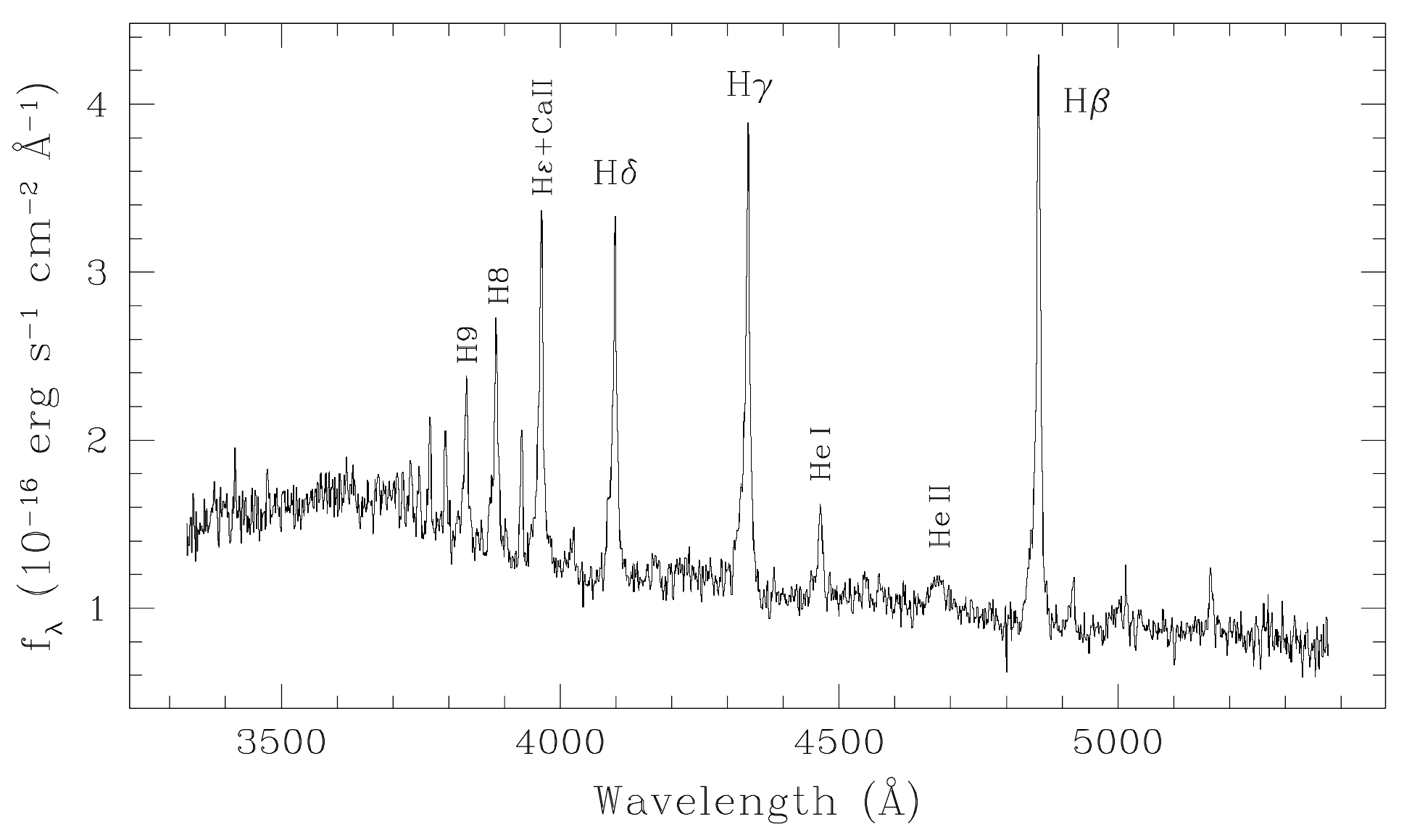}
\caption{First WHT/ISIS-blue spectrum of KR\,Aur taken on the night of 2008 December 16. The WD absorption lines are likely outshone by emission produced in an accretion event. Note the presence of the \hel{ii}{4686} emission line. Two-pixel FWHM Gaussian smoothing has been applied to the spectrum.}
\label{fig_20081216_first_sp}
\end{figure}

In order to investigate whether the flare events are associated with accretion episodes we computed integrated flux ratios between KR\,Aur and the reference star on the slit (neglecting any slit losses) over the 4000--5000 (blue arm) and 6500--7500\,\AA\ (red arm) wavelength ranges in the 2008 December 16 WHT/ISIS spectra. The resulting blue and red light curves are presented in Fig.~\ref{fig_flare_specphot}. The first spectrum of this series showed no sign of the WD absorption features and was dominated by strong Balmer and weaker \he{i} emission lines with noticeable \hel{ii}{4686} emission (Fig.~\ref{fig_20081216_first_sp}). The emission-line profiles exhibit enhanced wings extending to about $-1800$\,\kms\ in the blue and $600$\,\kms\ in the red. Subsequent spectra progressively revealed the WD photospheric absorptions with the emission lines becoming significantly narrower and less intense. The spectrophotometric light curves in Fig.~\ref{fig_flare_specphot} show a flux maximum at the start of the run followed by a decay, which suggests that KR\,Aur was experiencing a flare when these observations started. This together with the emission-line behaviour may indicate that the flare was actually an accretion event. Stronger emission lines with enhanced wings are also a feature of flare M-dwarf stars. However, \hel{ii}{4686} emission is rarely observed in these flare stars \citep[see e.g.][]{abraninetal98-1}. Furthermore, in the low state of the nova-like variable BB\,Dor the enhanced wings of the \Ha\ emission line produced a radial velocity curve  
with maximum blue excursion at orbital phase $\simeq 0.45$ and a delay of $\simeq 0.18$ cycle relative to the motion of the WD. These are characteristic features of the SW Sextantis stars and the magnetic AM Her stars in the high, accretion-driven state \citep[][and references therein]{rodriguez-giletal12-1}. This points to an accretion-related origin of the flare events, which show a wide range of durations.

\subsection{A magnetic accretion event?}
\label{sec:mag}

\begin{figure}
\centering
\includegraphics[width=\linewidth]{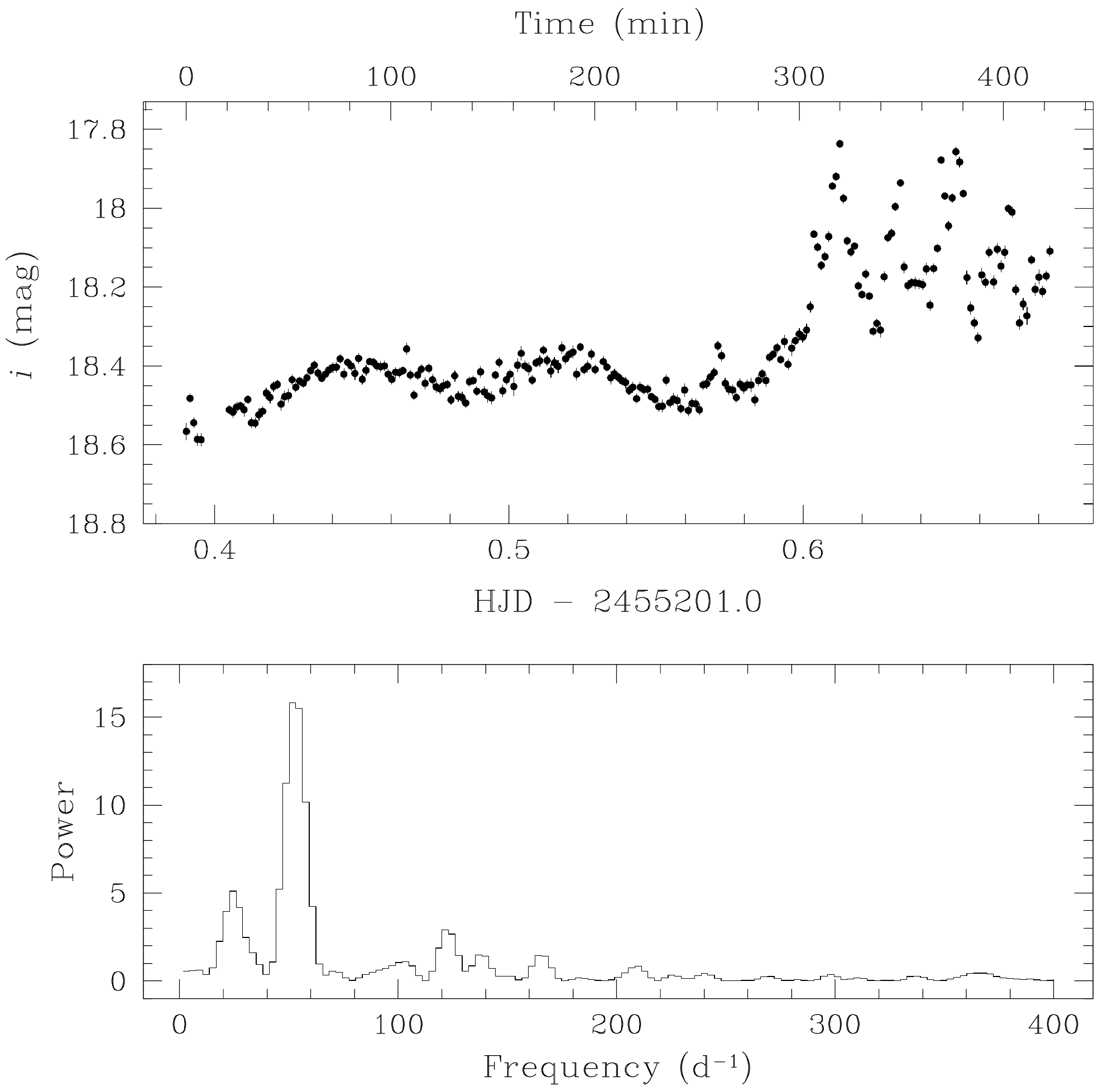}
\caption{Top panel: INT/WFC $i$-band light curve of KR\,Aur in the low state obtained on 2010 January 4. It shows the orbital ellipsoidal modulation with rapid brightness variations towards the end. The top $x$-axis is the time in minutes relative to the first data point of the light curve. Bottom panel: Scargle periodogram computed from the portion of this light curve lying in the abscissa interval between $\mathrm{HJD} - 2455201.0 = 0.601$ and 0.684.}
\label{fig:periodogram}
\end{figure}

On 2010 January 4 we observed an episode of rapid variability superimposed on the ellipsoidal light curve with a duration of at least two hours, preceded by a much weaker flare that occurred about 45 min earlier (Fig.~\ref{fig:periodogram}). As far as we know, this is the first detection of such behaviour in a nova-like variable in the low state. Unfortunately, on that night we had to stop the observation due to the onset of morning twilight, so the time coverage of these variations is limited. We computed a Scargle periodogram \citep{scargle82-1} from the data in the time interval between 0.601 and 0.684 ($\mathrm{HJD} - 2455201.0$), that we show in Fig.~\ref{fig:periodogram}. The dominant peak is centred at 27.4 min.

\begin{figure}
\centering
\includegraphics[width=\linewidth]{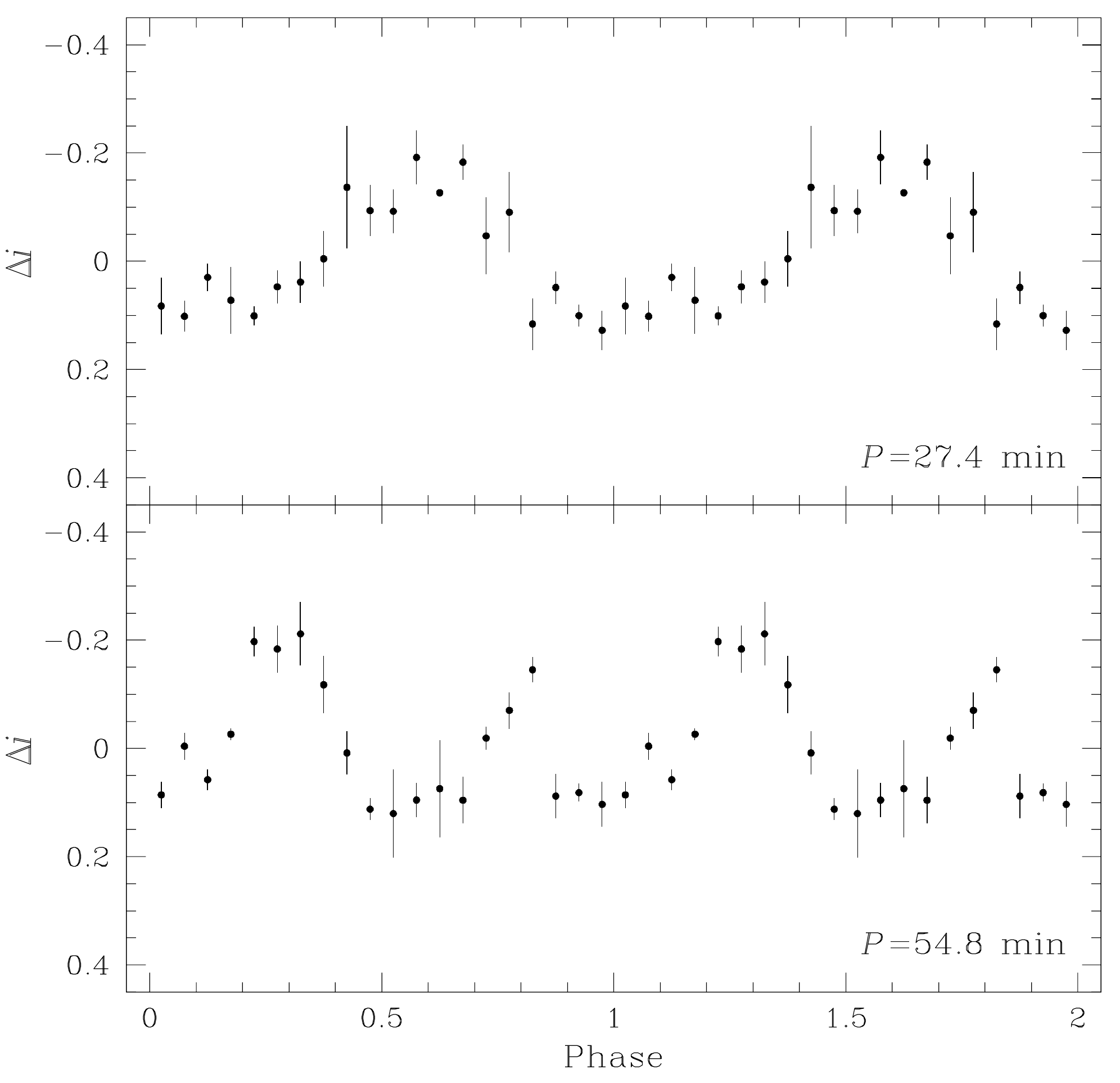}
\caption{Top panel: Light curve of the pulsating flare observed on 2010 January 4 folded on a 27.4-min period. The average-subtracted data points were phase-binned into 20 phase intervals and plotted twice for the sake of clarity. Bottom panel: same as above for a 54.8-min period.}
\label{fig:fold}
\end{figure}

Quasi-periodic pulsations (QPP) with time scales of tens of minutes have been observed in M dwarfs during flares. For example, a 32-min oscillation was detected in the M4.5 dwarf YZ CMi during the decay phase of a very intense ($-6$-mag amplitude) flare \citep{anfinogentovetal13-1}. Its $U$-band light curve showed a largely damped pulsation similar to longitudinal oscillations in solar flares. The maximum amplitude of the pulsating flare in KR\,Aur is about $-0.5$ mag, much smaller than the 6-mag flare in YZ CMi. One might think that the period of a QPP can increase with flare maximum energy, but this is not actually the case \citep{pughetal16-1}. Therefore, a QPP originated on the companion star can not be discarded. However, and despite the insufficient time coverage of the event observed in KR\,Aur on 2010 January 4, this pulsating flare does not exhibit the typical morphology of a stellar flare with a steep impulsive stage followed by a decay phase where QPP are damped, which may point to a non-QPP nature. 

\begin{table}
\centering
\caption{Times of ellipsoidal light curve minima at orbital phases 0.0 ($T_0$) and 0.5 ($T_{0.5}$)}
\begin{tabular}{@{}lccc}
\hline
Cycle & $T_{0}$/$T_{0.5}$ & $\sigma$ & Data \\
& (HJD -- 2450000) & (day) & \\
\hline
0.5      &   4874.4701  &  0.0012   & INT/WFC \\  
2009.5  &   5201.4803    &  0.0012   & INT/WFC \\
2230.5  &   5237.4540    &  0.0011   & INT/WFC \\
6420.5  &   5919.4648  &  0.0018   & IAC80/CAMELOT \\
11169.0  &   6692.3858   &  0.0015   & IAC80/CAMELOT \\
11169.5  &   6692.4604  &  0.0015   & IAC80/CAMELOT \\
13351.0  &   7047.5543   &  0.0016   & IAC80/CAMELOT \\
13357.0  &   7048.5310   &  0.0016   & IAC80/CAMELOT \\
13357.5  &   7048.6113   &  0.0016   & IAC80/CAMELOT \\
13362.5  &   7049.4247   &  0.0018   & IAC80/CAMELOT \\
13363.0  &   7049.5076   &  0.0018   & IAC80/CAMELOT \\
17897.0  &   7787.5098   &  0.0016   & IAC80/CAMELOT \\
17995.0  &   7803.4663   &  0.0005   & IAC80/CAMELOT \\
\hline
\end{tabular}
\label{tab:pha0}
\end{table}

Our alternative possible explanation for the observed oscillations is emission from one or two spots close to the poles of a magnetic WD right after accretion of the available material. Although our limited time coverage demands taking any interpretation with caution, pulses with two different amplitudes seem to alternate. This might point to a double-pulse modulation at twice the period of the dominant peak (i.e. 54.8 min). In Fig.~\ref{fig:fold} we show this rapid oscillation folded on 27.4 and 54.8 min after averaging the data points into 20 phase bins. The occurrence of alternating brighter and fainter pulses may suggest a double-humped oscillation on the spin period of the WD in KR\,Aur. Such modulations with a dominant first harmonic of the spin frequency have been observed in intermediate-polar CVs such as YY Dra \citep{pattersonetal92-1}, PQ Gem \citep{hellieretal94-2,potteretal97-1}, and V405 Aur \citep{allanetal96-1}. In KR\,Aur this scenario would require a mass transfer event and subsequent magnetic channelling of the material onto the WD. \cite{rodriguez-giletal12-1} and \cite{schmidtobreicketal12-1} have linked the enhanced wings of the emission lines and the appearance of two satellite emission S-waves in \Ha\ in the low-state spectra of the nova-like variable BB\,Dor with events of sporadic magnetic accretion of material supplied by a magnetically-active donor star, a behaviour also observed in the AM Her stars.

In the tentative magnetic accretion scenario the spin period of the WD in KR\,Aur would be either 27.4 or 54.8 min (with two pulses per spin cycle in the latter case), approximately. Note that the 27.4-min period that we detect in the low-state pulsating flare lies very close to the 25-min modulation observed by \cite{biryukov+borisov90-1} in the $B$-band light curve of KR\,Aur in the high state. However, as mentioned in Section~\ref{sec:intro} this 25-min modulation was observed in a light curve spanning less than four hours and has never been recovered.    

\subsection{Photometric ephemeris of KR Aur}
\label{sec_porb}

The only two attempts at determining the orbital period of KR\,Aur from radial velocity studies are those of \cite{shafter83-1} and \cite{hutchingsetal83-1}, with respective values of $0.16280 \pm 0.00003$ and $0.16274 \pm 0.00003$ day. We will derive the times of minima in our set of $i$-band ellipsoidal light curves to obtain a more accurate value of the orbital period and an ephemeris. The minimum light times are presented in Table~\ref{tab:pha0}. These were computed by fitting a Gaussian function to each minimum after masking out any flares or short-term activity. As mentioned in Section~\ref{sec:pblc} we assumed that the deeper minimum of the ellipsoidal light curve of KR\,Aur actually takes place at orbital phase 0, an assumption later validated in Section~\ref{sec:RVCs}.

The long time baseline spanned by our photometry and the limited accuracy of the previous determinations of the orbital period suggest an iterative approach in order to minimise uncertainties in the cycle count when deriving a new value of the orbital period. Therefore, we first computed a linear ephemeris using the orbital period reported by \cite{hutchingsetal83-1} and the mid-times of two phase-0.5 minima close in time to each other observed on 2010 January 4 and February 9 (second and third line in Table~\ref{tab:pha0}). Using Hutchings et al.'s orbital period and its uncertainty these two minima are found to be 221 orbital cycles apart with a cumulative error of 0.04 cycle. We then included the phase-0.5 minimum on 2009 February 11 (first line in Table~\ref{tab:pha0}) and used this initial linear ephemeris to compute the cycles between this and the 2010 January 4 minimum, which amount to 2009 with a cumulative error of 0.09 cycle. A further iteration including the mid-time of the phase-0.5 minimum on 2011 December 23 provided an orbital period accurate enough to provide a reliable cycle count for all the minima mid-time values in Table~\ref{tab:pha0}. The final ephemeris for KR\,Aur from a weighted linear fit is:    
\begin{equation}
T_0(\mathrm{HJD}) = 2454874.38975(68) + 0.162771641(49) \times E~,
\label{eq:ephem0}
\end{equation}
where numbers between brackets indicate the uncertainties in the last digits. All orbital phases in this paper were calculated using this ephemeris.

\begin{figure}
\centering
\includegraphics[width=\linewidth]{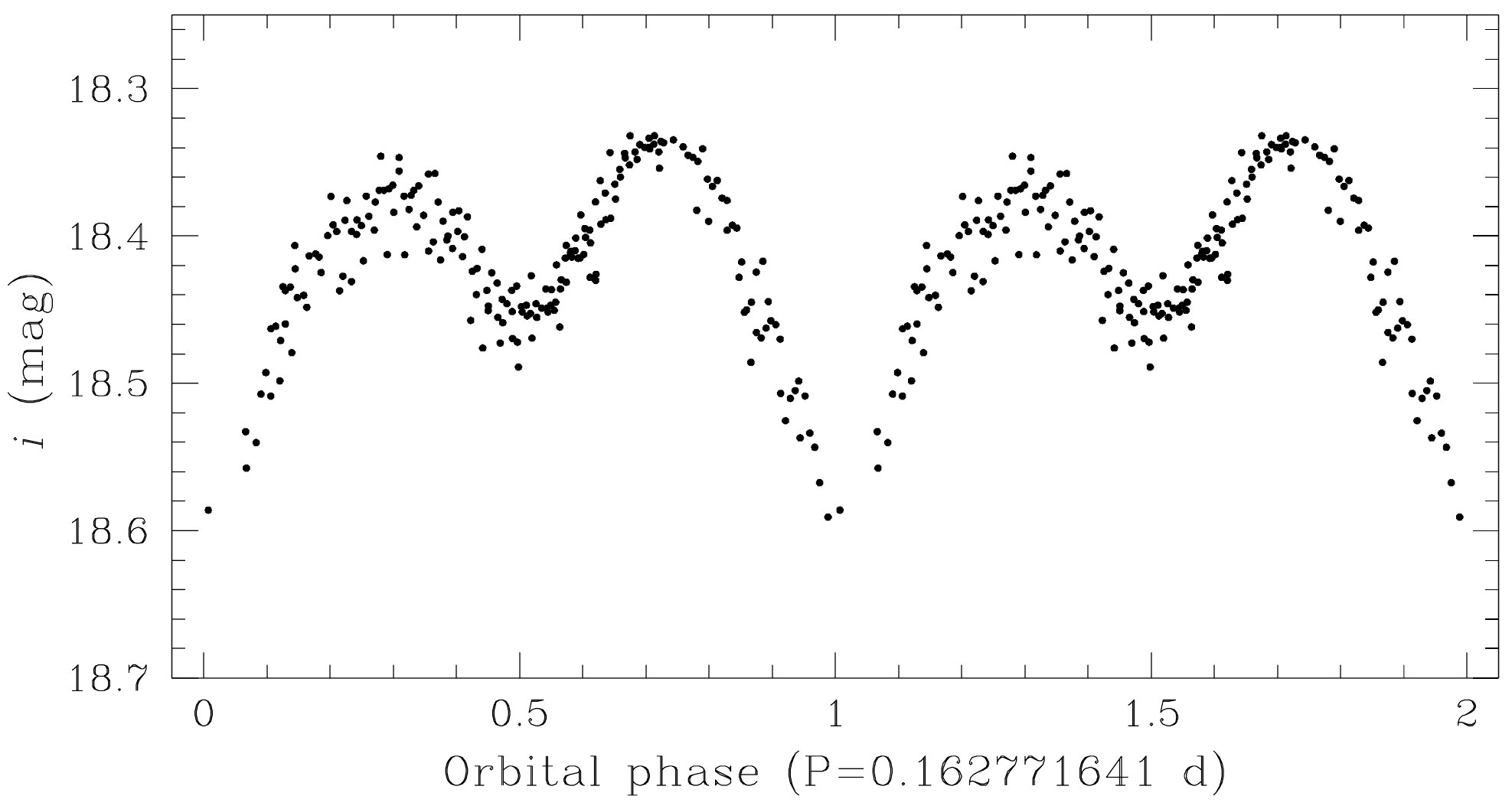}
\caption{Phase-folded $i$-band light curve of KR\,Aur in the low state. The phase-0.5 minimum is shallower than the phase-0 one due to irradiation of the inner hemisphere of the companion star by the WD. In computing this average light curve we only used the light curves that showed weak short-term activity superimposed on the ellipsoidal variation, and any flares were masked out. An orbital cycle has been repeated for the sake of clarity.}
\label{fig_folded_lc}
\end{figure}

In Fig.~\ref{fig_folded_lc} we present the phase-folded $i$-band light curve of KR\,Aur in the low state. The ellipsoidal modulation produced by the varying projected area of the donor star can be clearly seen. The folded light curve shows that the maximum at phase 0.25 is fainter than its phase 0.75 counterpart. In addition, the scatter observed around the former may indicate long-term brightness variability due to starspots \citep[e.g.][see also Section~\ref{sec:binparam}]{torresetal14-1}.

%%% SPECTRAL CLASSIFICATION %%%

\section{Spectral classification}
\label{sec:spec_class}

\begin{figure*}
\centering
\includegraphics[width=0.8\linewidth]{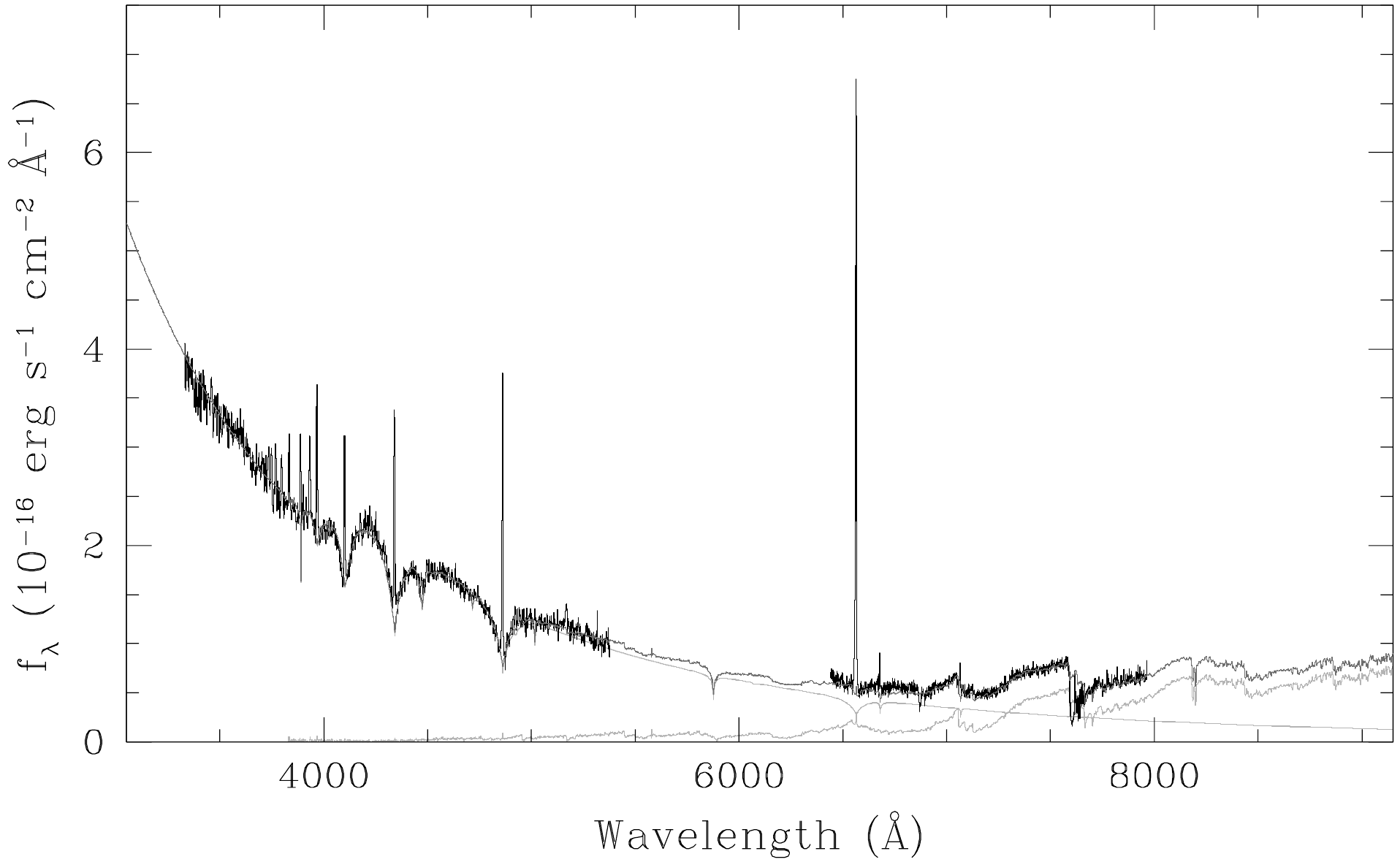}
\caption{Average WHT/ISIS spectrum of KR\,Aur in the low state (black). The best-fit composite model is shown in dark grey. It includes a DAB WD with $T_{1}=27\,148$\,K, $\log\,g=8.90$, and $\log (\mathrm{He/H})= -0.79$, and an M4.5\,V dwarf, both plotted in light grey.}
\label{fig:sed}
\end{figure*}

For spectral modelling we only considered the WHT/ISIS spectra taken out of flare events (spectra number 5 to 10; orbital phase between $0.69 - 0.28$; see Fig.~\ref{fig_flare_specphot}. Later spectra were discarded due to slit flux losses in the blue part of the spectrum).  We also neglected the GTC data due to dubious instrumental response correction in the redder part. In Fig.~\ref{fig:sed} we present the average WHT/ISIS spectrum of KR\,Aur in the low state that we used for the modelling.

The low-state blue spectrum of KR\,Aur is dominated by the characteristic broad Balmer absorptions produced on the WD photosphere. Superimposed on the absorptions are narrow emission components (intrinsic FWHM $\approx 240$ km s$^{-1}$ at \Ha). We will show in Section~\ref{sec:RVCs} that these are produced on the companion star. The spectrum also shows \he{i} absorption lines also with narrow emission lines on top and the \Ion{Ca}{ii} emission triplet in the near infrared. The presence of hydrogen and neutral helium absorption features indicates a DAB-type WD in KR\,Aur. This is the second detection of a DAB WD in a cataclysmic variable after HS\,0220+0603 \citep{rodriguez-giletal15-1}. In the red part of the spectrum the distinctive molecular bands of a late M-dwarf star are apparent. 

To compute the atmospheric parameters ($T_1$ and $\log g$) of the WD and its relative helium abundance we compared the WD blue spectrum with a grid of synthetic DAB WD spectra \citep{koester10-1}. The DAB model grid we used spans effective temperatures $T_{1}=10\,000-30\,000$\,K in steps of $200$\,K, surface gravities $\log\,g=7.0-9.8$ in steps of 0.2 dex, and helium abundances $\log\,(\mathrm{He/H})$ between $-5.0$ and $0.0$ in steps of 0.2 dex. We first normalised both the average and model spectra to the continuum and masked the narrow Balmer and \Ion{He}{i} emissions. The mask widths were selected so as to completely neglect these narrow emission lines. An additional 2 and 4~\AA\ were masked on both sides of the \Ion{He}{i} and \Ion{H}{i} emission lines, respectively, to prevent distortion in the residual absorption lines. As a starting point of this analysis we used $\chi^2$ minimisation to constrain the parameter space. 

\begin{figure}
\centering
\includegraphics[width=8.3cm]{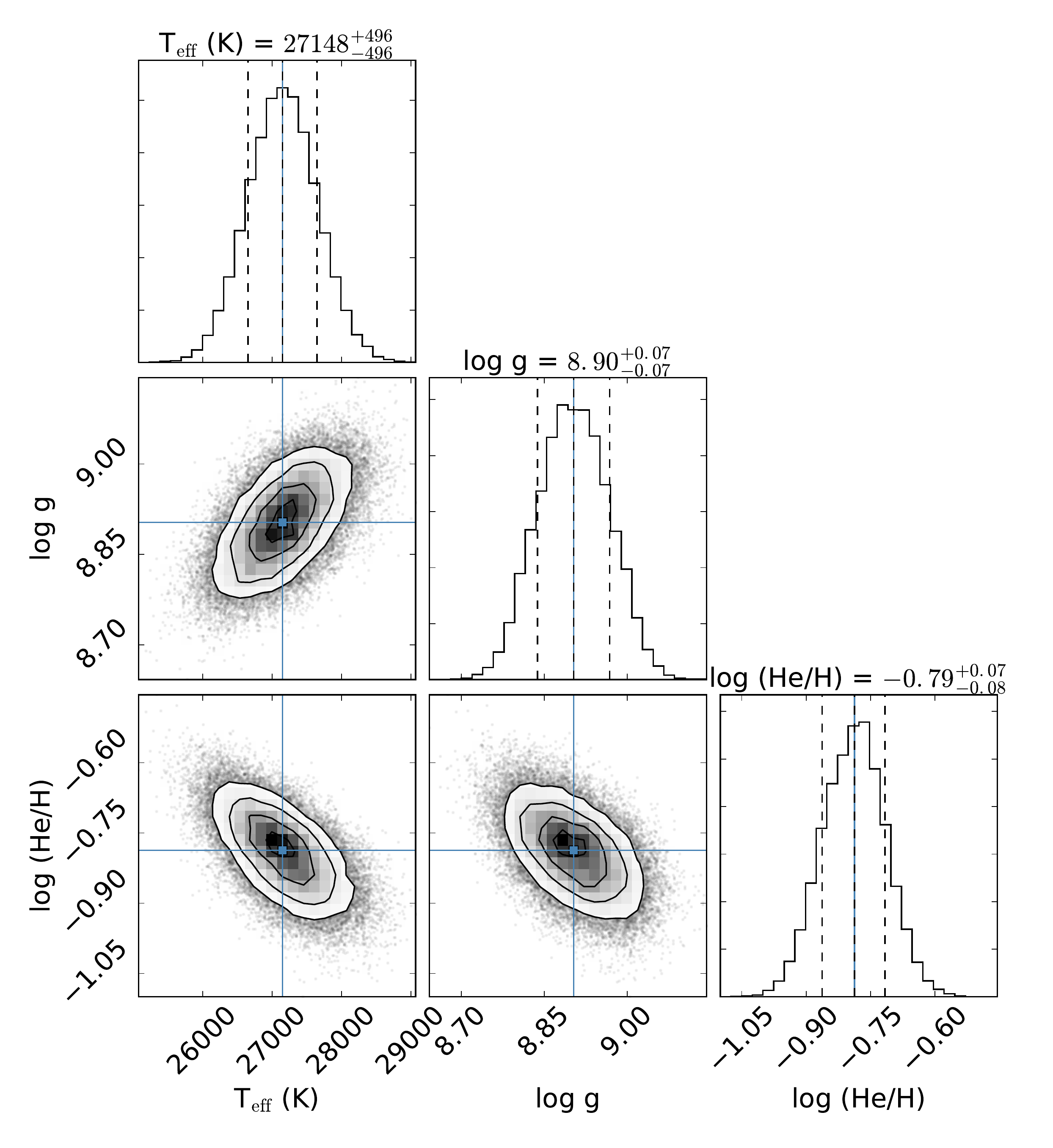}
\caption{Correlation diagrams for the probability distribution of the best-fit WD atmospheric parameters:  $T_{1}=27\,148 \pm 496$\,K, $\log g=8.90 \pm 0.07$, and $\log (\mathrm{He/H})= -0.79$\,$_{-0.08}^{+0.07}$.}
\label{fig:corner_plot_WD}
\end{figure}

Next, we carried out the spectral fitting with the Markov-chain Monte Carlo (MCMC) package {\sc emcee} for Python \citep{foreman-mackeyetal13-1}, constraining the three parameters with flat priors based on the previous $\chi^2$ minimisation: $T_{1}=25\,000-29\,000$\,K, $\log\,g=8.6-9.8$, and $\log\,(\mathrm{He/H})=-2.0-0.0$. For the MCMC simulation we used 30 random seeds (walkers) as input for the chains with 80\,000 iterations per seed. This produced the best-fit WD parameters $T_{1}=27\,148$\,$\pm$\,$496$\,K, $\log\,g=8.90 \pm 0.07$, and $\log (\mathrm{He/H})= -0.79$\,$_{-0.08}^{+0.07}$ (quoted uncertainties are 1$\sigma$; see Fig.~\ref{fig:corner_plot_WD}). 

\begin{figure}
\centering
\includegraphics[width=\linewidth]{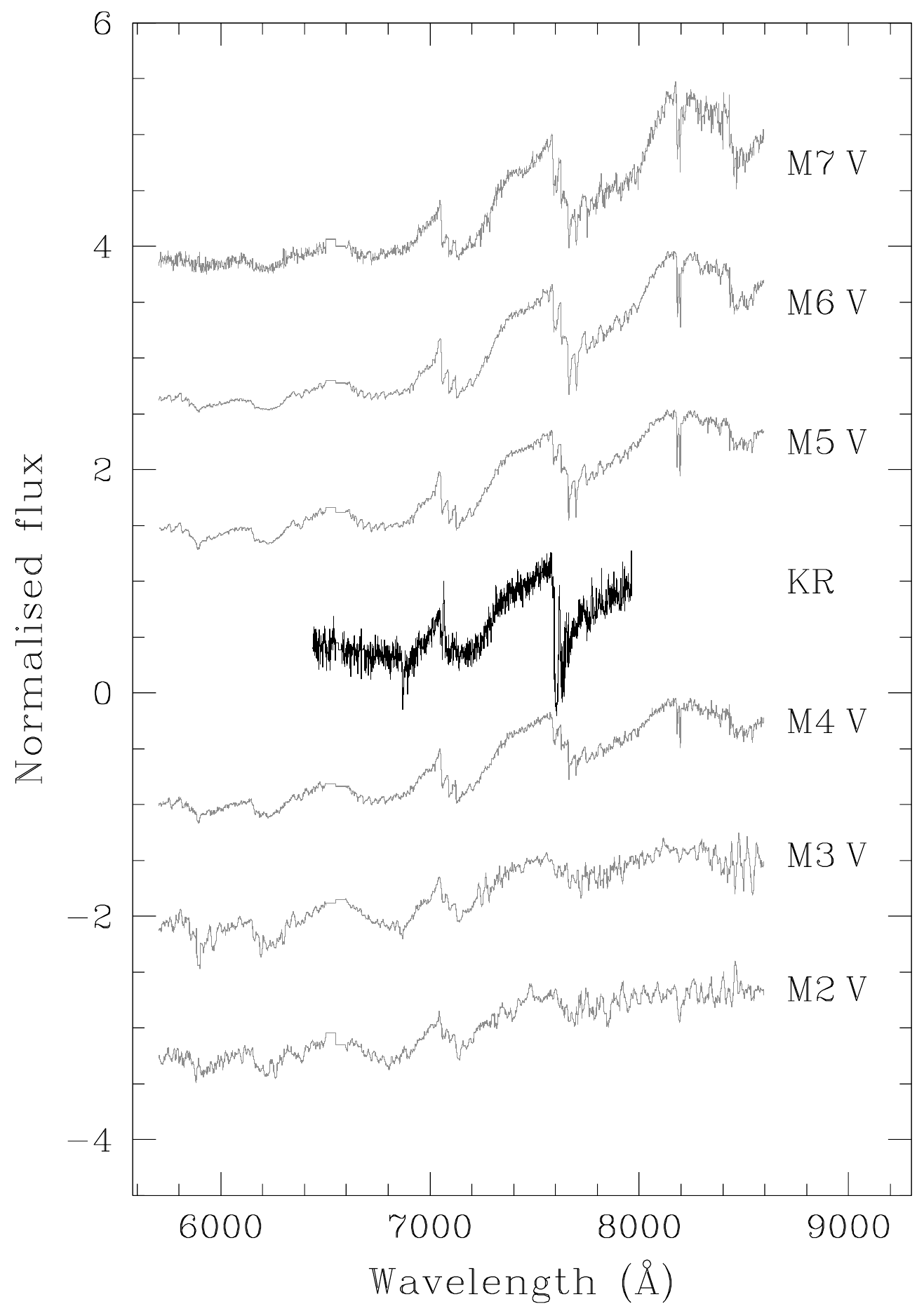}
\caption{Comparison of the spectrum of the M-dwarf companion in KR\,Aur with spectra from a SDSS library of red dwarfs. Note that the best-fit WD model spectrum was previously subtracted from the target spectrum. All spectra were normalised to their flux value at 7500 \AA. Offsets in steps of 1.25 have been applied for the sake of visualisation.}
\label{fig_mdwarf_comparis}
\end{figure}

In order to obtain the spectral type of the companion star we first subtracted the best-fit WD model from the average WHT/ISIS red data to remove the WD continuum contribution. After masking out the emission and telluric absorption lines from the WD-subtracted average spectrum, we compared it with a library of Sloan Digital Sky Survey (SDSS) M-dwarf spectra covering the spectral types M0--M9\,V in steps of one subtype after normalising to the flux value at 7500 \AA. Minimisation of the $\chi^2$ gives a best-fit spectral type of M4.5\,V. Using this best-fit spectral type we determined an $i$-band fractional contribution of the companion star to the total observed flux of $f = 0.56$ (i.e. a veiling factor $f_\mathrm{veil} = 1 - f = 0.44$). A comparison of the average red spectrum of KR\,Aur with spectra of M dwarfs of different types is presented in Fig.~\ref{fig_mdwarf_comparis}. By visual inspection we can safely rule out M3 or earlier spectral types. Further improvement on the classification of the M dwarf would require a larger spectral range targeting specific absorption molecular bandheads \citep[see][and references therein]{rodriguez-giletal15-1}.

The best-fit atmospheric parameters of the WD can be used to determine its mass and radius, and here we will compare the results delivered by three mass-radius relations. The sequences for carbon/oxygen cores by \citet*{fontaineetal01-1} yield a WD mass $M_1=1.16 \pm 0.03$\,\Msun\ and radius $R_1=0.0063 \pm 0.0004$\,\Rsun\ (thick, $M_{\mathrm{H}}/M_1=10^{-4}$ hydrogen layer, pure-hydrogen atmosphere), and $M_1=1.15 \pm 0.03$\,\Msun, $R_1=0.0063 \pm 0.0004$\,\Rsun\ (thin, $M_{\mathrm{H}}/M_1=10^{-10}$ hydrogen layer, pure-helium atmosphere). Using the mass-radius relation of \citet{camisassaetal19-1} for ultra-massive WDs ($1.1 - 1.29$\,\Msun) with oxygen/neon cores and hydrogen-rich atmospheres we obtain $M_1=1.14 \pm 0.03$\,\Msun, $R_1=0.0063 \pm 0.0003$\,\Rsun. In Section~\ref{sec:binparam} we will compare these masses with the dynamical mass that we have obtained for the WD in \target.

\begin{figure}
\centering
\includegraphics[width=\linewidth]{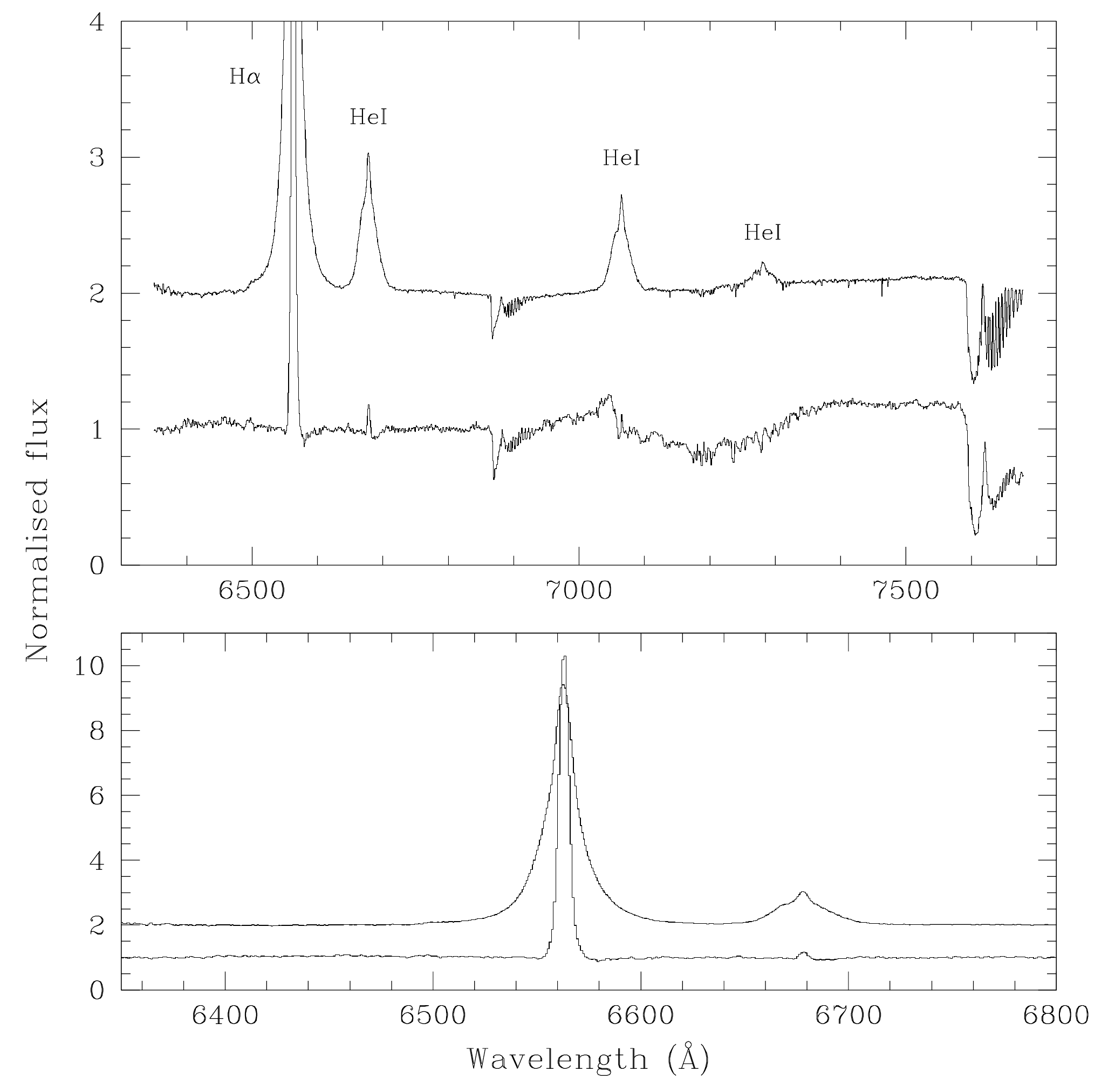}
\caption{Top panel: Comparison between the average spectrum of KR\,Aur in the low state (GTC/OSIRIS; bottom spectrum) and in the intermediate state (Gemini/GMOS; top spectrum, offset by 1 in the ordinate axis). Bottom panel: Zoomed-in version of the top panel in the \Ha+\hel{i}{6678} region to illustrate the large change in line width due to ongoing accretion in the intermediate state.}
\label{fig:redav}
\end{figure}

\section{Comparison of low and intermediate state spectra}
\label{sec:interm}

We measured an $i$-band magnitude of 17.1 for KR\,Aur on the GMOS acquisition image. Thus, during the Gemini visit the system was about 1.3 mag brighter than it was in the low state, when the GTC and WHT observations were performed. We will hereafter refer to this brighter state as the intermediate state. The Balmer and \he{i} emission lines in all the GMOS spectra obtained in the intermediate state (4.6-h continuous coverage) are much broader, with extended wings, and more intense than they are in the low state (see Fig.~\ref{fig:redav}). This is a clear indication of ongoing accretion in the system at a lower level than in the high state.

\begin{figure*}
\centering
\includegraphics[width=\linewidth]{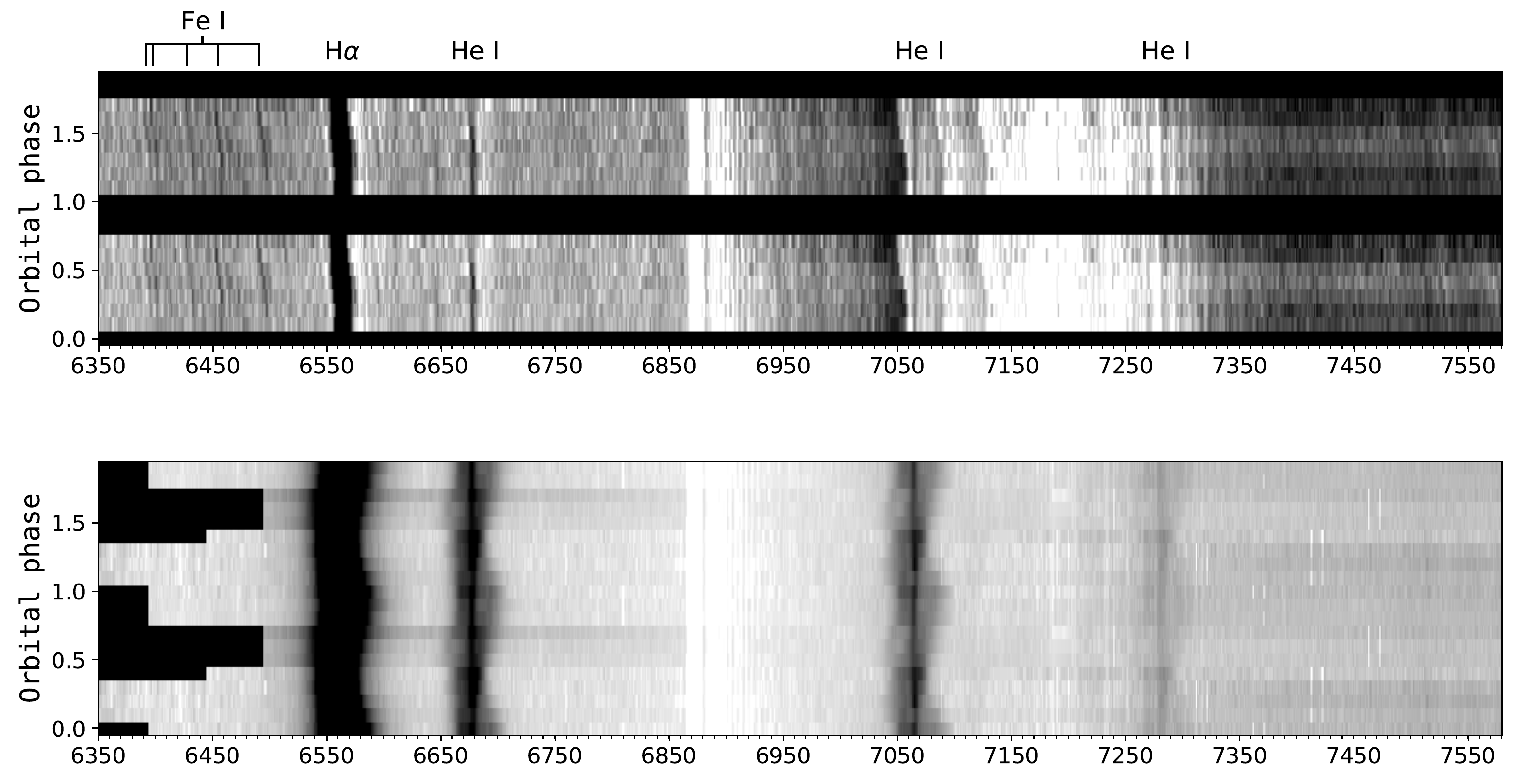}\\
\includegraphics[width=\linewidth]{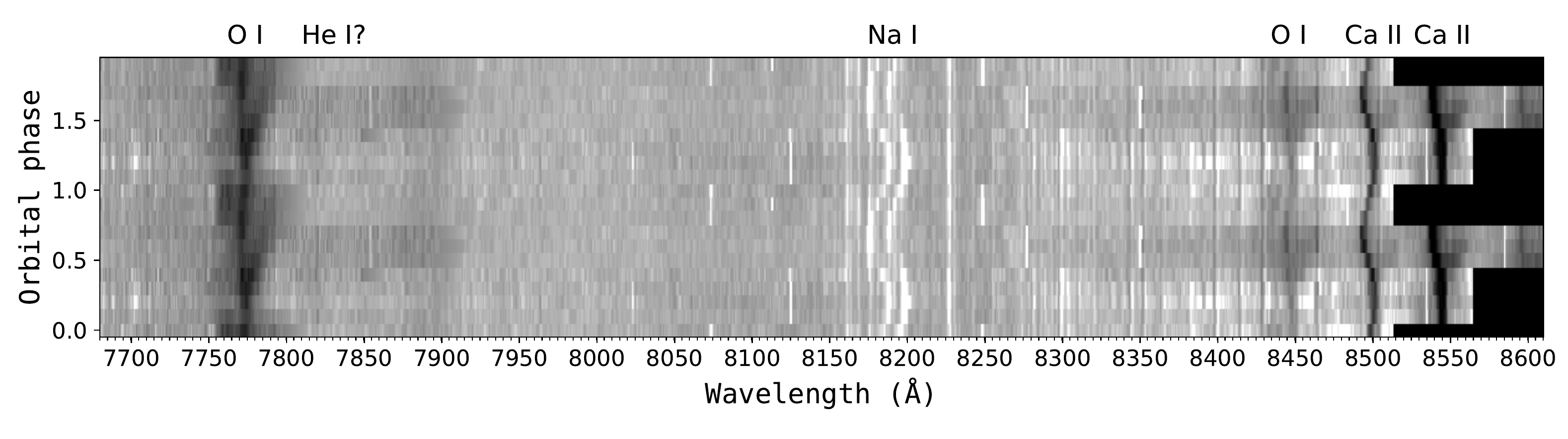}
\caption{Trailed spectra greyscale plots of KR\,Aur built by phase-folding the continuum-normalised spectra into 10 bins using the photometric ephemeris (Eq.~\ref{eq:ephem0}). Top panel: low-state GTC/OSIRIS data. Middle and bottom panels: intermediate-state Gemini/GMOS data. The most prominent spectral features are identified (the top and middle panels share the same line identification labels except for the \Ion{Fe}{i} emission lines observed in the low state). The orbital cycle has been repeated for the sake of clarity. The black stripes are either non-sampled orbital phases (low state) or due to differences in wavelength coverage (intermediate state; caused by dithering in the spectral direction to fill in the GMOS detector gaps).}
\label{fig:trailed}
\end{figure*}

In Fig.~\ref{fig:trailed} we show the low-state (GTC) and intermediate-state (Gemini) trailed spectra. In the low state the \Ha\ and \Ion{He}{i}, as well as a forest of \Ion{Fe}{i} emission lines, follow the orbital motion of the companion star. The latter show maximum flux around phase 0.5, a sign of irradiation from the WD. \Line{O}{i}{7773} emission from the companion is also present. Other features originating on the companion star are observed in the 7050--7140\,\AA\ wavelength range, which we identify as TiO. In the trailed spectrogram their appearance in emission is an artefact due to uncertain continuum normalisation of the regions dominated by the absorption bands originating on the M-dwarf companion. 

\begin{figure*}
\includegraphics[width=12cm,center]{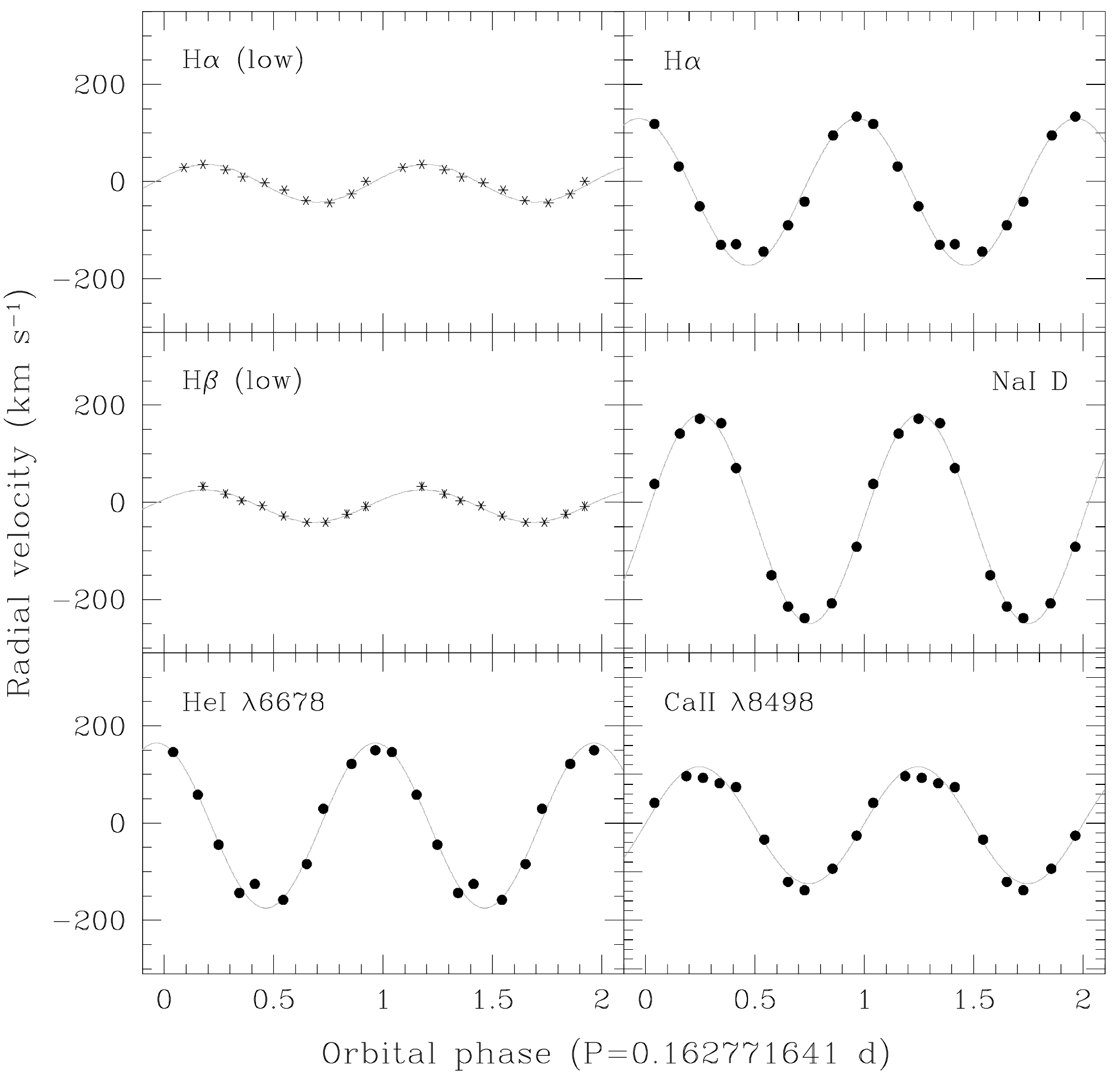}
\caption{Radial velocity curves of selected spectral lines. The lines labelled ``low'' were measured on spectra obtained in the low state, while the remainder were measured on intermediate-state spectra. We phase-binned the data into 10 bins and plotted the orbital cycle twice for the sake of clarity. The solid lines are the best sine fits to the data.} 
\label{fig:RVCs}
\end{figure*}

In the intermediate state the line profiles show more complexity: broader emission S-waves with maximum blue shift at approximately phase 0.5 can be clearly seen together with narrow emission cores in \Ha, \Ion{He}{i}, and \Ion{O}{i} $\lambda$7773, $\lambda$8446. These components produce "swordfish"-like emission-line profiles in the average spectrum (see Fig.~\ref{fig:redav}). The \Ion{Ca}{ii} triplet is dominated by narrow emission from the companion star, and the \Ion{Na}{i} absorption doublet is also apparent.
 
\section{Radial velocity curves}
\label{sec:RVCs}

Radial velocity curves of the emission lines in the low state were derived by cross-correlation of the profiles with single Gaussian templates. The template FWHMs were computed from Gaussian fits to the average line profiles. For the intermediate-state spectra we used the double-Gaussian technique of \citet{schneider+young80-2} with a Gaussian separation of 2000\,\kms\ to probe the motion of the emission-line wings. In both cases we first re-binned the spectra to achieve a constant velocity step centred on the rest wavelength of each line and phase-binned the spectra into 10 bins using the photometric ephemeris given in Eq.~\ref{eq:ephem0}. We fit the resulting radial velocity curves with sine functions of the form:
\begin{equation}
V = \gamma + K \sin[2\pi(\varphi-\varphi_0)]\,,
\end{equation}
where $\gamma$ is the radial velocity offset, $K$ the radial velocity amplitude, $\varphi$ the orbital phase, and $\varphi_0$ the phase offset. The resulting best-fit parameters are presented in Table~\ref{tab:RVCs}.

In the low-state spectra we only analysed \Ha\ and \Hb. The other emission lines (e.g. \he{i}) are too faint to be feasibly fit. However, in the intermediate state the emission lines are much stronger and wider, so we were able to derive the radial velocity curves of some \he{i} lines and \Line{Ca}{ii}{8498}. For the latter we used the single Gaussian approach fixing the FWHM of the template at 250\,\kms. In measuring the radial velocity of the \Line{Na}{i}{8183, 8195} absorption doublet we used a different strategy. We first computed a preliminary radial velocity curve of the absorption component at 8183.27\,\AA\ by fitting Gaussian functions. We then removed the orbital velocity variation from the spectra by subtracting the best sine fit to that initial radial velocity curve. A template absorption doublet was subsequently created by averaging the velocity-corrected profiles. We finally obtained the radial velocity curve of the \Ion{Na}{i} absorption doublet by cross-correlation with this template. 

The \Ion{Na}{i} radial velocity curve has a null phase offset with respect to the photometric ephemeris computed in Section~\ref{sec_porb} (Eq.~\ref{eq:ephem0}). The fact that this doublet originates on the companion star confirms our zero phase assumption, i.e. the deeper minimum in the ellipsoidal light curve (Fig.~\ref{fig_folded_lc}) actually takes place at zero phase, contrary to what is expected for a non-irradiated companion. The radial velocity curves of the Balmer emission lines in the low state and the \Line{Ca}{ii}{8498} emission line in the intermediate state also indicate an origin on the companion side of the binary system for these lines.

The radial velocity curves measured from the \Ha, \hel{i}{6678} and \hel{i}{7065} emission-line wings in the intermediate state show a 0.72-cycle offset with respect to the motion of the companion star, hence producing maximum blue shift at orbital phase 0.47. Relative to the expected motion of the WD (i.e. maximum blue shift at $\varphi = 0.25$) this corresponds to a delay of 0.22 cycle. A similar delay of 0.18 cycle was found in the radial velocity curve of the \Ha\ emission-line wings of the nova-like variable BB Dor during accretion episodes in the low state \citep{rodriguez-giletal12-1}. This phase offset relative to the radial velocity curve of the WD is a defining characteristic of the SW Sextantis stars \textit{in the high state}, when mass transfer from the companion is fully developed at a large rate. These same offsets are seen in the high states of AM Her stars (highly-magnetic CVs). In AM Her stars in the high state the broad emission components are thought to arise at different parts of the mass stream from the companion star including the region when the stream couples with the magnetic field lines of the WD \citep[see e.g.][]{schwopeetal97-1,simicetal98-1}. Our expectation is that the radial velocity curve of the emission-line wings will have maximum blue excursion at approximately orbital phase 0.45, similarly to the accretion events detected in BB Dor. We have shown that KR\,Aur also displays accretion episodes during the low state. However, only few such events were recorded, thus preventing a radial velocity study of the emission-line wings. 

\begin{table*}
\caption{Best sine fit parameters of the radial velocity curves. $\varphi_0$ is the phase offset relative to the ephemeris given by Eq.~\ref{eq:ephem0}.}
\centering
\begin{tabular}{@{}lccc}
\hline
Line & $\gamma$ & $K$ & $\varphi_{0}$\\
& (\kms) & (\kms) & \\
\hline
{\textbf{Low state}} & & & \\
\Ion{Fe}{i} forest & $-9.2 \pm 6.9$ & $143.2 \pm 6.0$ & $0.007 \pm 0.011$ \\
\Ha & $-3.61 \pm 0.08$ & $39.0 \pm 0.1$ & $-0.0548 \pm 0.0005$ \\
\Hb & $-7.6 \pm 2.8$ & $33.4 \pm 3.4$ & $-0.07 \pm 0.02$ \\
\hline
{\textbf{Intermediate state}}  &  &  &  \\
\Ha\ wings & $-21.4 \pm 0.3$ & $151.3 \pm 0.5$ & $0.7191 \pm 0.0005$ \\
\hel{i}{6678} wings & $-5.0 \pm 1.2$ & $169.8 \pm 1.7$ & $0.715 \pm 0.002$ \\
\hel{i}{7065} wings & $-64.5 \pm 1.5$ & $142.0 \pm 2.2$ & $0.723 \pm 0.002$ \\
\Line{O}{i}{7773} wings & $8.9 \pm 8.3$ & $251.0 \pm 9.9$ & $0.699 \pm 0.007$ \\
\Line{Na}{i}{8183, 8195} & $-34.4 \pm 2.1$  & $215.1 \pm 2.6$ & $0.0003 \pm 0.0026$ \\
\Line{Ca}{ii}{8498} & $-4.4 \pm 0.7$ & $120.2 \pm 1.1$ & $-0.004 \pm 0.001$ \\
\hline
\end{tabular}
\label{tab:RVCs}
\end{table*}

%%% BINARY PARAMETERS %%%

\section{Binary parameters}
\label{sec:binparam}

In order to determine the fundamental parameters of KR\,Aur we simultaneously modelled
the $i$-band light curve and the \Ion{Na}{i} absorption-line and \Ion{Ca}{ii} emission-line radial velocity curves using  \textsc{xrbcurve}. All these data were obtained when \target\ was in the low state. The model and methods used are fully described in \citeauthor{shahbazetal00-1} (\citeyear{shahbazetal00-1}, \citeyear{shahbazetal03-1}, \citeyear{shahbazetal17-1}), and its application to an eclipsing nova-like variable is reported in \citet{rodriguez-giletal15-1}. Briefly, the model includes the effects of heating of the companion star by a point
source from the compact object. We use \textsc{nextgen} model-atmosphere fluxes
\citep*{hauschildtetal99-1} to determine the intensity distribution on the companion
star and a quadratic law to correct the intensity for
limb-darkening \citep{claret+bloemen11-1}.  For the radial velocity curves we specify the
strength of the absorption or emission lines over the companion star's surface, and
integrate to obtain the corresponding line-of-sight radial velocity. We set the
line strength given by its equivalent width (EW) according to the effective
temperature for each element on the star.  Using \textsc{phoenix} model spectra
\citep{husseretal13-1} we determine the EW-temperature relationship for the
\Ion{Na}{i} (absorption) and \Ion{Ca}{ii} (emission) lines. The line strengths for each surface
element on the star are then calculated using its temperature and this
EW-temperature relation.  For the effect of external heating we introduce the
limiting factors $F_{\rm AV}$ and $F_{\rm EV}$ for the absorption- and
emission-line flux, respectively.  If the external radiation flux is greater
than some fraction of the unperturbed flux, then we set the EW for that element
to zero and so it does not contribute to the radial velocity curve.

In determining the binary parameters we use a MCMC 
method convolved with a differential evolution fitting algorithm  within a
Bayesian framework \citep[see][and references within]{shahbazetal17-1}.  We made use
of flat prior probability distributions for the model parameters.
We employ 20 individual chains to
explore the parameter space and 40\,000 iterations per chain. We reject the first
100 iterations and only include every second point.

As shown in Section~\ref{sec_porb} (see also Fig.~\ref{fig_folded_lc}) the ellipsoidal light curve of \target\ shows unequal maxima. The decrease in flux near phase
0.25 may be due to the presence of a dark star spot on the surface of
the companion star. Indeed, there is observational evidence of dark star spots
in binary systems (\citealt{hilletal14-1}; \citealt*{shahbazetal14-1}; \citealt{parsonsetal16-1}; \citealt{hilletal17-1}). These are due to strong magnetic
activity on the companion star. Therefore, we model the light and radial
velocity curves including a dark spot on the receding hemisphere 
of the star.

The model parameters that determine the shape and amplitude of the light  and
radial velocity curves are the mass of the WD, $M_{\rm 1}$, the mass of the companion star, $M_{\rm 2}$, the orbital inclination, $i$, the distance to the target, $D_{\rm pc}$, the unabsorbed heating flux, $\log F_{\rm X,0}$, and the effective temperature of the companion star, $T_{\rm 2}$. The other  parameters are the light curve
phase shift $\delta \phi_{\rm LC}$, the  absorption- and emission-line radial
velocity phase shifts, $\delta  \phi_{\rm AV}$ and $\delta \phi_{\rm EV}$,
respectively, and the absorption- and emission-line velocity offsets,
$\gamma_{\rm AV}$ and  $\gamma_{\rm EV}$,  respectively. Finally, the dark spot
parameters: position, $C_{\rm spot}$, width, $W_{\rm spot}$, normalisation,
$T_{\rm  spot}$, and extent, $E_{\rm spot}$. From our measured $f_{\rm veil}$ =
0.44 (see Section~\ref{sec:spec_class}) it is clear that there is an extra light component
that veils the observed light from the companion star. To allow for this veiling
we include an extra flux component, $f_{i}$, in the model light curve. 

The absorption- and emission-line radial velocity curves consist of 14 and 12 points, respectively. The $i$-band data points were phase folded according to the orbital ephemeris (Eq.~\ref{eq:ephem0}) and averaged into 42 orbital phase bins. We
assume a fully Roche-lobe-filling companion star and fix the gravity-darkening exponent to 0.08 for a convective star \citep{lucy67-1}. The use of intermediate-state radial velocity curves justifies the assumption of a companion that fully fills its Roche lobe. However, this assumption might not be satisfied by the $i$-band photometry that was obtained when \target\ was in the low state. We fix $\gamma_{\rm EV}$ to $\gamma_{\rm AV}$ because there is no reason why they should be different if they arise from the surface of the companion star. We also fix the colour excess\,$E(B-V) = 0.07$ (Section~\ref{sec:spec}). Using the results obtained in Section~\ref{sec:spec_class} we adopt a WD temperature $T_{\rm 1}$ = 27\,000 K and an effective temperature of the companion star $T_{\rm 2}=3100$\,K, right between an M4\,V and a M5\,V star \citep{bessell91-1}.

A preliminary fit shows that a dark spot on the receding hemisphere of the companion star is not sufficient to model the whole light curve. This model fails to match the minimum around phase 0.0 and $f_{\rm veil}$ does not match the observed value by a factor of $\simeq 2$. We also find that the fits are not sensitive to $F_{\rm AV}$, which means that the model requires that all the inner face of the star contributes to the radial
velocity curve. Therefore, we repeated the MCMC fitting procedure including two dark spots,
one on the receding hemisphere and one on the backside of the star. A quick look at the 1D parameter distributions revealed that, apart from $F_{\rm AV}$, most of the parameters were well constrained. Therefore, for speed and to reduce the number of model parameters we fixed the spot parameters to their optimal values of temperature ($T_{\rm spot}$) and fractional surface coverage of $T_{\rm spot} = 336$\,K and 0.084 (backside spot), and $T_{\rm spot} = 1867$\,K and 0.099 (receding hemisphere spot), and the limiting factor for the absorption-line flux to the value that the preliminary fit provided ($F_{\rm AV} = 3.02$), and repeated the MCMC fits. We obtained a limiting factor for the emission-line flux, $F_{\rm EV} = 1.15^{+0.32}_{-0.10}$. This value can be affected by an additional source such as a partial accretion disc, that will shadow parts of the companion star. Since the radial velocity curves used in the fit come from spectra in the intermediate state, when a partial disc may be present, the value of $F_{\rm EV}$ has to be regarded as a lower limit.

\begin{table}
\centering
\caption{Results obtained from the simultaneous model fit to the $i$-band optical light curve and the \Ion{Na}{i} absorption-line and \Ion{Ca}{ii} emission-line radial velocity curves. Quoted uncertainties are 1$\sigma$. The fit was conducted for an effective temperature of $T_2 = 3100$\,K, halfway between the M4\,V and M5\,V spectral types.}
\begin{tabular}{l l l }
\hline   
\\
         \multicolumn{2}{c}{--- Model parameters --- } \\
\\
$T_{2}$  (K)           & 3100\\ %& 3200   \\
\rule{0pt}{5ex}$M_{1}$  (\Msun)        & 0.94$^{+0.15}_{-0.11}$\\   %& 0.93$^{+0.16}_{-0.12}$ \\
\rule{0pt}{5ex}$M_{2}$  (\Msun)        & 0.37$^{+0.07}_{-0.07}$\\   %& 0.36$^{+0.09}_{-0.07}$ \\
\rule{0pt}{5ex}$\cos\,i$              & 0.68$^{+0.03}_{-0.01}$\\   %& 0.68$^{+0.04}_{-0.02}$ \\
\rule{0pt}{5ex}$D_{\rm pc}$ (pc)      & 546$^{+41}_{-38}$\\        %& 603$^{+52}_{-50}$ \\
\rule{0pt}{5ex}$\log F_{\rm X,0}$ (\ergs)  & $-11.67^{+0.06}_{-0.06}$\\  %& $-11.71^{+0.07}_{-0.07}$ \\
\\
\\
	\multicolumn{2}{c}{--- Derived parameters --- } \\ 
\\
$q$                   &  0.39$^{+0.03}_{-0.04}$\\   %& 0.39$^{+0.04}_{-0.04}$ \\
\rule{0pt}{5ex}$K_{\rm 2}$ (\kms)     & 223.2$^{+4.8}_{-4.9}$\\    % & 223.1$^{+5.5}_{-5.3}$ \\
\rule{0pt}{5ex}$R_{\rm 2}$ (\Rsun)    & 0.41$^{+0.03}_{-0.03}$\\    %& 0.42$^{+0.03}_{-0.02}$ \\
\rule{0pt}{5ex}$v\sin\,i$  (\kms)     & 93.6$^{+5.8}_{-6.1}$\\      %& 94.3$^{+4.5}_{-4.4}$ \\
\rule{0pt}{5ex}$f$    &  0.53 \\                     %& 0.52 \\
\\
\hline                
\end{tabular}
\label{table:results}
\end{table}

%%%%%%%%%%%%%%%%%%%%%%%%%%%%%%%% 1 %%%%%%%%%%%%%%%%%%%%%%%%%%%%%%%%%
\begin{figure*}
  \begin{center}
\includegraphics[width=0.9\linewidth,angle=0]{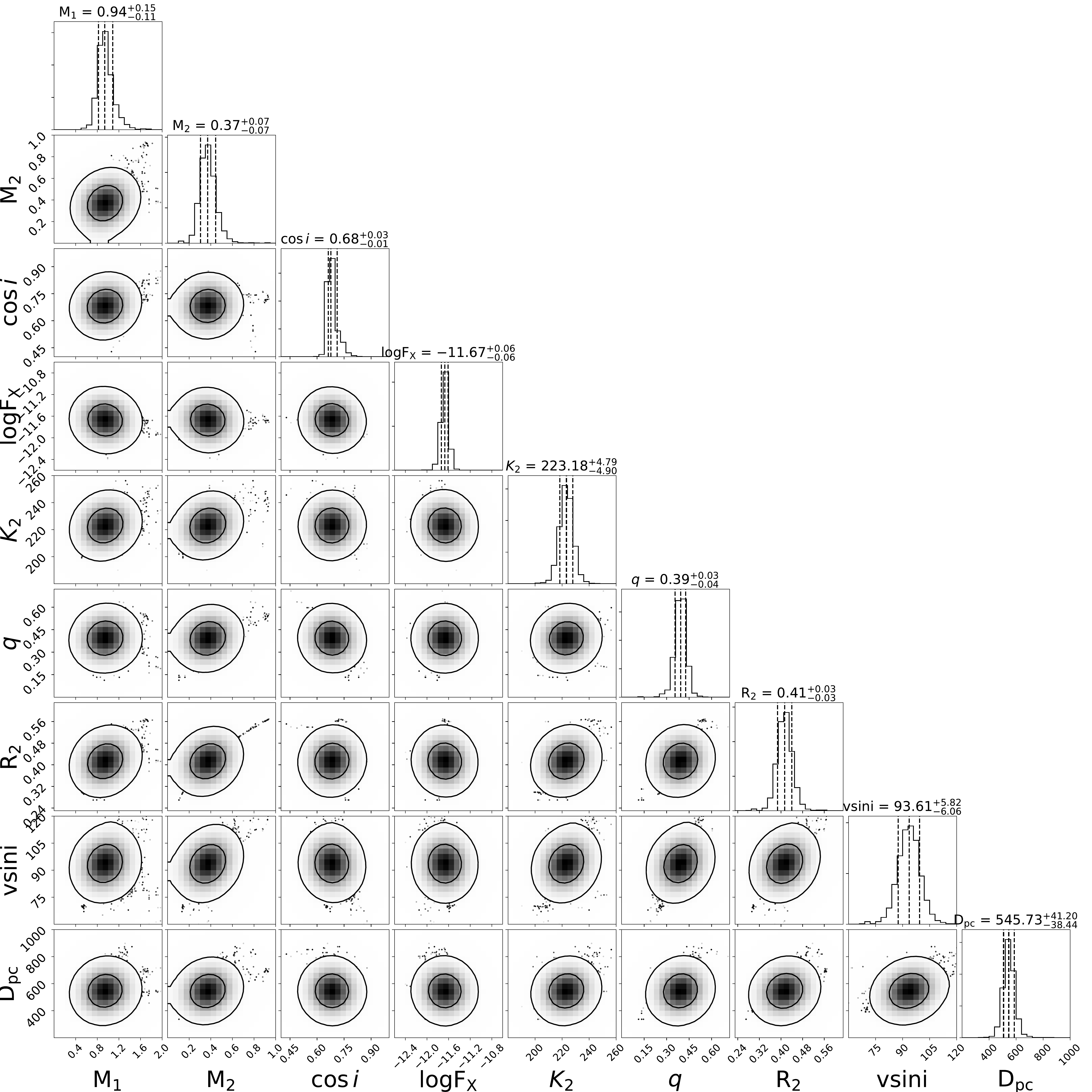}
\caption{
MCMC 2-D model parameter distributions resulting from the fits to the light and radial velocity curves of \target\ using our model with 
$T_{\rm 2}=3100$\,K. The contours in the 2-D plots show the 1$\sigma$, 2$\sigma$, and 3$\sigma$ confidence regions, and the right panels show the projected 1-D parameter distributions with the mean 
and standard deviation. The companion star's volume equivalent radius, $R_2$, and 
its projected rotational velocity, $v \sin i$, are inferred values derived from the radial velocity 
amplitude, $K_2$, and the binary mass ratio, $q$.
} 
\label{figure:MCMC}
 \end{center}
\end{figure*}

\begin{figure*}
  \begin{center}
\includegraphics[width=0.9\linewidth,angle=0]{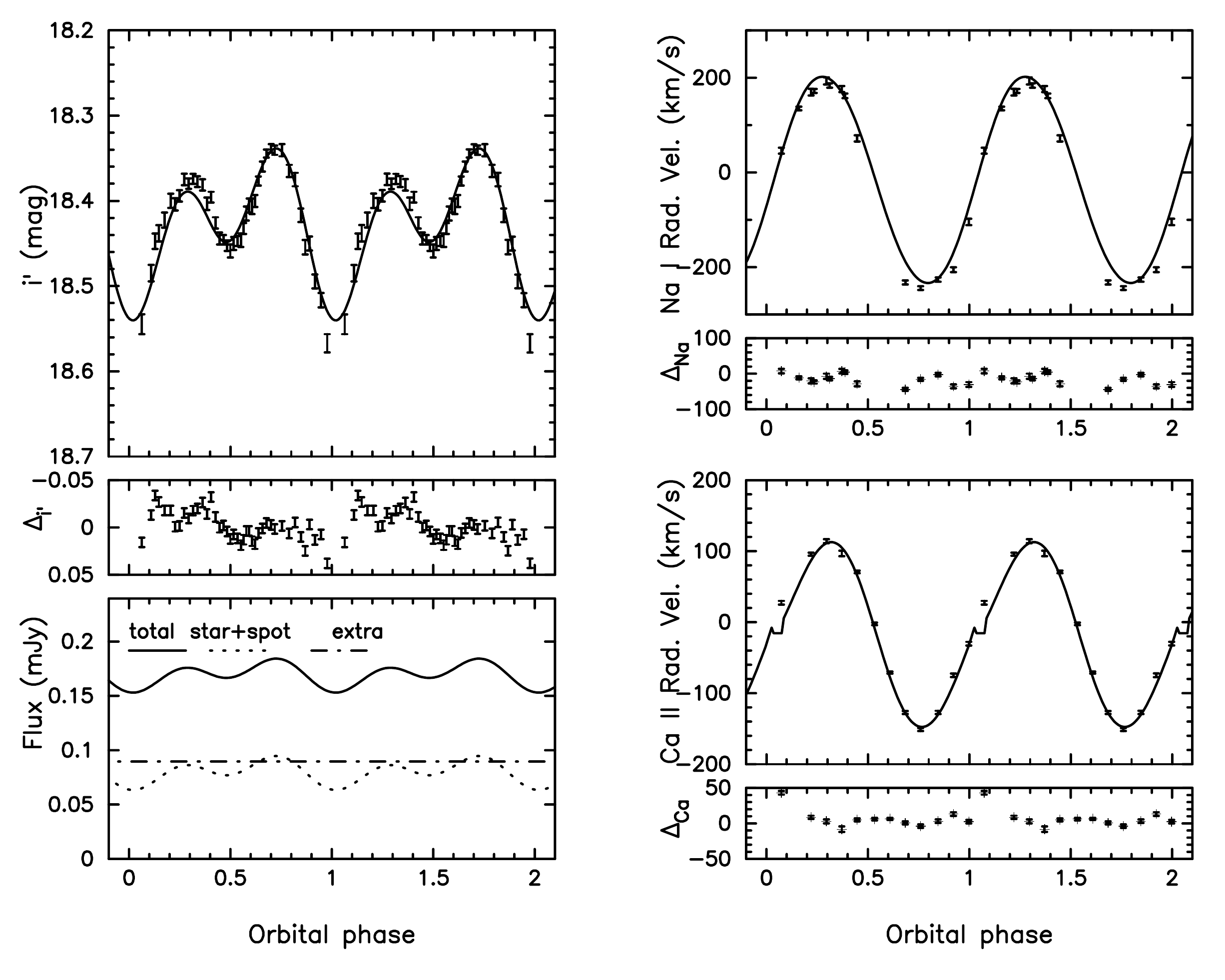}
\caption{
The results of the simultaneous fits to the light and radial velocity 
curves of \target\ using our model. The $i$-band light curve (top left), the 
\Ion{Na}{i} absorption-line (top right) and \Ion{Ca}{ii} 
emission-line (bottom right) radial velocity curves are shown along with 
the corresponding best fit and residuals. The bottom left 
plot shows the flux of the individual components in the system: the companion star plus dark spots (dashed line), extra light from the WD (dot dashed lines) and the total flux 
(black solid line). The orbital cycle has been repeated for the sake of clarity.
} 
\label{figure:bestmodelfit}
 \end{center}
\end{figure*}

The final set of parameters determined from our \textsc{xrbcurve} fits and their 1$\sigma$ uncertainties are given in Table~\ref{table:results}. The MCMC parameter distributions are shown in Fig.~\ref{figure:MCMC}. All model parameters are well constrained, and the overall agreement between the data and model is good. In Fig.~\ref{figure:bestmodelfit} we show the best simultaneous fit to the light- and radial-velocity curves. From the best fit we determine the fractional contribution of the companion star ($f=1-f_{\rm veil}$) in the $i$-band to be 0.53 for $T_2=3100$\,K, which is close to the 0.56 that we derived in Section~\ref{sec:spec_class} from the low-state spectra. The distance derived from the model is consistent within the uncertainties with the value from the \textit{Gaia} DR2 parallax, that ranges from 377 to 563\,pc \citep{bailer-jonesetal18-1}.

Our modelling yields a WD irradiating flux ($F_{\rm X,0}$) of approximately $2.1 \times 10^{-12}$\,erg~s$^{-1}$\,cm$^{-2}$. This translates into a WD luminosity of $\simeq 5 \times 10^{31}$\,erg~s$^{-1}$ at the \textit{Gaia} DR2 distance of 451\,pc. Using the effective temperature and the radius obtained for the WD in Section~\ref{sec:spec_class} we derive a luminosity of $\simeq 7 \times 10^{31}$\,erg\,s$^{-1}$. The agreement between these two values strengthen our assumption of the WD being responsible for the heating of the companion star. 

%%%%%%%%%%%%%%%%%%%%%%%%%%%%%%%%%%%%%%%%%%%%%%%%%%%%%%%%%%%%%%%%%%%%%%%%

%%%%%%%%%%%%%%%%%%%%%%%%%%%%%%%%%%%%%%%%%%%%%%%%%%%%%%%%%%%%%%%%%%

The two independent WD mass measurements determined from the dynamical analysis ($0.94\,\Msun$; Table~\ref{table:results}) and the spectral modelling ($\simeq 1.15\,\Msun$; Section~\ref{sec:spec_class}) disagree at the 1$\sigma$ level, with the spectroscopic mass estimate being larger than the dynamical one. We investigated whether orbital phase smearing of the Balmer line profiles could result in an overestimate of the spectroscopic WD mass, but found that the radial velocity amplitude of the WD ($K_1 = q\,K_2 \simeq 87$\,\kms) is too low to produce significant broadening of the absorption lines. However, the detailed choice of the wavelength ranges masking the emission lines in the cores of the Balmer absorption profiles does affect the spectroscopic fit. While increasing the size of the emission-line masks did not produce a significant change in the WD parameters, and hence its inferred mass, decreasing their size yielded larger surface gravities and therefore larger WD masses. We conclude that remaining contamination of the Balmer absorption lines by the emission lines from the companion star may result in a slight overestimate of the surface gravity. An alternative possibility is that the WD is weakly magnetic, $\lesssim 5$\,MG (see fig.\,8 in \citealt{rodriguez-giletal15-1} and e.g. \citealt{wickramasinghe+ferrario00-1}), as already suggested in Section~\ref{sec:mag}. The magnetic nature of the WD in \target\ could be probed with spectropolarimetry obtained during a low state, which allows to detect the Zeeman $\sigma^{+,-}$ components of the Balmer absorption lines even against the contamination of the emission lines from the companion star. However, note that the difference between the WD dynamical mass and its spectroscopic mass is well within 2$\sigma$, so evidence for a magnetic WD scenario is inconclusive based on these grounds alone.

Using eq.~2 of \cite{townsley+gaensicke09-1} with the effective temperature and the dynamical mass of the WD in \target\ in results in a time-averaged mass transfer rate ($\dot M$) of about $5.5 \times 10^{-10}$~\Msun~yr$^{-1}$, which is not excessively high compared to MV\,Lyr, TT\,Ari and DW\,UMa, and is closer to the derived value of $1.1 \times 10^{-9}$~\Msun~yr$^{-1}$ for HS\,0220+0603 \citep{rodriguez-giletal15-1}. This suggests that CVs within the 3--4\,h period range may exhibit a large scatter in their average accretion rates, which contrasts the earlier predictions of a narrow and well-defined $\dot M$ vs $P$ relation \citep[see e.g. fig.\,3 of][]{howelletal01-1}. The possibility of a wide spread in accretion rates is supported by more recent CV population models that include a wider range of evolution channels  \citep[see e.g. fig.\,1 of][]{goliasch+nelson15-1}. In addition, a large scatter in WD effective temperatures has been observed in CVs above the orbital period gap \citep{palaetal17-1}.

Only one dynamical mass determination of a nova-like variable in the 3--4 hour interval existed previous to the present work. \citet{rodriguez-giletal15-1} obtained the fundamental parameters of the eclipsing SW Sextantis star HS\,0220+0603 from observations in the low state. Their main results were the finding of the first DAB WD in a CV with a mass of 0.87\,\Msun\ and an effective temperature of $\simeq 30\,000$\,K, and a M5.5\,V companion star (measured around zero orbital phase) too cool for its mass of 0.47~\Msun. The same mismatch is observed in \target: the derived dynamical mass of the companion star in \target\ is 0.37~\Msun, which is too large for a spectral type around M5\,V \citep[predicted mass $0.12-0.16$\,\Msun;][]{kniggeetal11-1}.

%%% CONCLUSIONS %%%

\section{Conclusions}
\label{sec:conclu}

We have conducted a dynamical study of the non-eclipsing nova-like variable \target\ in the low state. From a simultaneous fit to the $i$-band ellipsoidal light curve, the \Ion{Na}{i} absorption doublet and  \Ion{Ca}{ii} emission triplet radial velocity curves we have found stellar masses of $M_1=0.94^{+0.15}_{-0.11}$\,\Msun\ and $M_2=0.37^{+0.07}_{-0.07}$\,\Msun\ for the WD and the companion star, respectively, and an orbital inclination of approximately $47^\circ$. In the fitting process we had to include two dark spots on the companion star, one on its receding hemisphere and one on its backside, to be able to reproduce the unequal maxima observed in the $i$-band ellipsoidal light curve. The distance obtained from the model is consistent with the \textit{Gaia} DR2 distance $451^{+112}_{-75}$\,pc, making \target\ a nearby nova-like variable.   

We refined the orbital period of \target\ to $0.16277164 \pm 0.00000005$\,day, or $3.906519 \pm 0.000001$\,h, and derived an orbital ephemeris from our set of $i$-band ellipsoidal light curves.

We found a best-fit spectral type of the companion star of M4.5\,V, which is too cool for its mass as also observed in the eclipsing nova-like variable HS\,0220+0603 in the low state.

Comparison of the average WD spectrum of \target\ with a grid of synthetic spectra yielded a DAB type with an effective temperature $T_1=27\,148 \pm 496$\,K, surface gravity $\log g=8.90 \pm 0.07$, and a relative helium abundance $\log (\mathrm{He/H})= -0.79_{-0.08}^{+0.07}$. The WD mass derived from these parameters and mass-radius relations is $M_1 \simeq 1.15$\,\Msun. We suggest that this larger mass relative to the dynamical mass might be due to contamination by the narrow emission lines originated on the companion star or by extra broadening of the Balmer absorption lines produced by the magnetic field of the WD. 

Related to this, we found a sudden photometric oscillation in one of our $i$-band light curves with a period of either 27.4 or 54.8\,min that we tentatively associate with the spin of the WD being revealed by a magnetic accretion event. Therefore, we suggest that the WD in \target\ may be magnetic.

The light curve of \target\ in the low state shows relatively frequent flares of $\approx - 0.5$\,mag from the underlying level. Spectrophotometry on one night revealed that these flare events may likely be triggered by accretion episodes. At event peak the absorption lines of the WD are no longer visible and the spectrum gets dominated by strong Balmer and weaker \Ion{He}{i} emission lines with presence of \Ion{He}{ii} emission, a characteristic trait of accretion.

We have also presented the analysis of time-resolved spectra when \target\ was in an intermediate brightness state about 1.3 mag brighter than it was in the low state. The emission lines show a high-velocity emission S-wave with the 0.2-cycle delay relative to the motion of the WD typical of the SW Sextantis stars in high state, indicating a re-establishing of accretion at a rate lower than in maximum brightness.

%%% ACKNOWLEDGEMENTS %%%

\section*{Acknowledgements}

We thank the anonymous reviewer for a prompt and useful report. PR-G acknowledges support from the State Research Agency (AEI) of the Spanish Ministry of Science, Innovation and Universities (MCIU) and the European Regional Development Fund (FEDER) under grant AYA2017--83383--P. MAPT acknowledges support from the State Research Agency (AEI) of the Spanish Ministry of Science, Innovation and Universities (MCIU) under grant AYA2017--83216--P. He also acknowledges support by a Ram\'on y Cajal Fellowship (RYC--2015--17854). BTG and OT were supported by the UK STFC grant ST/P000495. OT was also supported by a Leverhulme Trust Research Project Grant. We are grateful to Nicolas Lodieu and Marcela Espinoza for providing their library of M-dwarf spectra. Our thanks also go to Carlos Allende Prieto for useful discussion. We thank all the M--1 group observers for the long-term dedication to the search for low states in nova-like variables. We also thank the Editor in Chief of \textit{Astronomicheskii Tsirkulyar}, Dr Georgij M. Rudnitskij, for kindly providing the text of \cite{biryukov+borisov90-1}. We acknowledge with thanks the variable star observations from the AAVSO International Database contributed by observers worldwide and used in this research. This research has made use of the APASS database, located at the AAVSO web site. Funding for APASS has been provided by the Robert Martin Ayers Sciences Fund. This research has made use of the ``Aladin sky atlas'' developed at CDS, Strasbourg Observatory, France. The use of the {\sc pamela} and {\sc molly} packages developed by Tom Marsh is acknowledged. Based on observations made with the Gran Telescopio Canarias (GTC), installed at the Spanish Observatorio del Roque de los Muchachos of the Ins\-ti\-tu\-to de Astrof\'\i sica de Canarias (IAC), on the island of La Palma. This article is also based on observations made at the Observatorios de Canarias of the IAC with the William Herschel Telescope (WHT) and the Isaac Newton Telescope (INT), operated on the island of La Palma by the Isaac Newton Group in the Observatorio del Roque de los Muchachos, and on observations obtained at the Gemini Observatory (Program ID GN--2008B--Q--39) which is operated by the Association of  Universities for Research in Astronomy, Inc., under a cooperative agreement with the NSF on behalf of the Gemini partnership: the National Science Foundation (United States), the National Research Council (Canada), CONICYT (Chile), the Australian Research Council (Australia), Minist\'{e}rio da Ci\^{e}ncia, Tecnologia e Inova\c{c}\~{a}o (Brazil) and Ministerio de Ciencia, Tecnolog\'\i a e Innovaci\'on Productiva (Argentina). Also based on observations obtained with the SARA Observatory 1.0 m telescope at the Observatorio del Roque de los Muchachos, which is owned and operated by the Southeastern Association for Research in Astronomy (\url{saraobservatory.org}).

%%%%%%%%%%%%%%%%%%%%%%%%%%%%%%%%%%%%%%%%%%%%%%%%%%



\footnotesize{

\bibliographystyle{mnras}
\bibliography{mn-jour,aabib}
%\bibliography{aabib}
}
%%%%%%%%%%%%%%%%%%%%%%%%%%%%%%%%%%%%%%%%%%%%%%%%%%

%%%%%%%%%%%%%%%%% APPENDICES %%%%%%%%%%%%%%%%%%%%%

%\appendix



%%%%%%%%%%%%%%%%%%%%%%%%%%%%%%%%%%%%%%%%%%%%%%%%%%

% Don't change these lines
\bsp	% typesetting comment
\label{lastpage}
\end{document}